\documentclass[journal]{IEEEtran}

\pdfoutput = 1

\ifCLASSINFOpdf
\else
   \usepackage[dvips]{graphicx}
\fi

\usepackage{cite}
\usepackage{url}
\usepackage{fancybox}
\usepackage{graphicx}
\usepackage{amsmath}
\usepackage{algorithmic}
\usepackage{algorithm}
\usepackage{booktabs}
\usepackage{epstopdf}
\usepackage{array}
\usepackage{fancybox}
\usepackage{amssymb}
\usepackage{bm}
\usepackage{makecell}
\usepackage{xcolor}
\usepackage{stfloats}

\begin{document}

\title{Structure Parameter Optimized Kernel Based Online Prediction with a Generalized Optimization Strategy for Nonstationary Time Series}

\author{Jinhua Guo,~\IEEEmembership{Student Member, IEEE,} Hao Chen, Jingxin Zhang, Sheng Chen,~\IEEEmembership{Fellow, IEEE} %
\thanks{J. Guo and H. Chen are with Fujian Provincial Key Laboratory of Intelligent Identification and Control of Complex Dynamic Systems, Fujian Institute of Research on the Structure of Matter, Chinese Academy of Sciences, Fuzhou 350002, China (E-mails: 18202237256@163.com, chenhao@fjirsm.ac.cn).} %
\thanks{J. Zhang is with Department of Automation, Tsinghua University, Beijing 100084, China (E-mail: zjx18@mails.tsinghua.edu.cn).} %
\thanks{S. Chen is with School of Electronics and Computer Science, University of Southampton, Southampton SO17 1BJ, U.K. (E-mail: sqc@ecs.soton.ac.uk).} %
\vspace*{-5mm}
} 

\maketitle

\begin{abstract}
In this paper, sparsification techniques aided online prediction algorithms in a reproducing kernel Hilbert space are studied for nonstationary time series. The online prediction algorithms as usual consist of the selection of kernel structure parameters and the kernel weight vector updating. For structure parameters, the kernel dictionary is selected by some sparsification techniques with online selective modeling criteria, and moreover the kernel covariance matrix is intermittently optimized in the light of the covariance matrix adaptation evolution strategy (CMA-ES). Optimizing the real symmetric covariance matrix can not only improve the kernel structure's flexibility by the cross relatedness of the input variables, but also partly alleviate the prediction uncertainty caused by the kernel dictionary selection for nonstationary time series. In order to sufficiently capture the underlying dynamic characteristics in prediction-error time series, a generalized optimization strategy is designed to construct the kernel dictionary sequentially in multiple kernel connection modes. The generalized optimization strategy provides a more self-contained way to construct the entire kernel connections, which enhances the ability to adaptively track the changing dynamic characteristics. Numerical simulations have demonstrated that the proposed approach has superior prediction performance for nonstationary time series. 
\end{abstract}


\IEEEpeerreviewmaketitle

\section{Introduction}\label{sec:1}

Online prediction of nonstationary time series is a particularly challenging and pervasive problem in many fields of signal processing, machine learning and process control. Conventional approaches and their parameters updating mechanisms may be insufficient to track the changing underlying dynamic characteristics due to nonstationarity \cite{hong2008model,8291833}. In the past several decades, kernel based online approaches in a reproducing kernel Hilbert space have been extensively studied, since the problem may become easier to solve with the so-called ``kernel trick'' if the observed input data are mapped onto a high-dimensional Hilbert space \cite{vapnik1995}. According to the representer theorem \cite{sch2001a}, the optimal solution to the minimization of the regularized empirical risk can be expressed as weighted kernels that composed of all the available training samples. However, in an online learning scenario, the dimension of the estimated weight vector will be continuously increasing along with the sequentially arrived data, which not only brings intractable computational complexity issue but also may degrade the generalization ability and hence decrease prediction performance \cite{7120172,1315946}. Therefore, in kernel based online modeling, the sparsification procedure is essential.

The sparsification techniques for online kernel modeling can be implemented in a supervised or unsupervised way, to properly select kernel dictionary. The unsupervised sparsification techniques construct appropriate kernel dictionary by just using the observed input data based on some selective modeling criteria, such as the approximate linear dependency (ALD) \cite{1315946,SANTOS2017114}, the coherence criterion \cite{4685707}, the distance criterion \cite{6515200,7553552}, and their combinations \cite{8291833,9147041}. The supervised sparsification techniques adopt a variety of criteria that also require the observed output data, such as the two-part novelty condition of novelty criterion \cite{6796766}, the subjective information measure of surprise criterion \cite{5337958}, the changed significance of loss function \cite{FAN20132174}, and the orthogonal forward selection (OFS) procedure of online tunable gradient radial basis function (RBF) networks \cite{9040439}. A particular sparsification approach is based on the subset selection of the training samples, which can also be viewed as an indirect approach to constructing the kernel dictionary, such as the information-theoretic learning (ITL) criterion \cite{paiva2017,FAN20144349}. Within the aforementioned sparsification techniques, there are two representative approaches for online prediction. One is the kernel adaptive filter (KAF) approach \cite{liu2010kernel}, and the other one is the online tunable RBF approach with the fixed kernel dictionary size~\cite{6334482,7310878}.

In addition to the kernel dictionary, the other structure parameter of online kernel modeling is the kernel covariance matrix in each kernel, which takes effect in the metric space that is formed by the observed input data. Without loss of generality, the Gaussian kernel is used as the default one in this paper. To prevent from inducing intractable computational tasks, most Gaussian kernel based online modeling approaches choose to use restricted forms of kernel covariance matrix, including the isotropic matrix, which is proportional to the identity matrix, and the diagonal matrix, which can only capture axis-aligned dynamic characteristics \cite{1315946,9040439,adaptivekernelsize}. Comparing to the restricted forms, the general symmetric covariance matrix considers the cross relatedness of input variables and it greatly improves the kernel structure's flexibility as well as the online prediction modeling performance at the expense of significantly increased computational complexity in optimizing the kernel covariance matrix. The general symmetric kernel covariance matrix can be interpreted as a coordinate transformation of shifting and rotating with respect to the original coordinates~\cite{bishop2006pattern,DBLP:journals/corr/Hansen16a}. 

The covariance matrix adaptation evolution strategy (CMA-ES) \cite{DBLP:journals/corr/Hansen16a} can be adopted to optimize the kernel covariance matrix for online modeling in the following two ways. Firstly, the evolving mechanism for its covariance matrix of the Gaussian distribution in the CMA-ES can be adopted or extracted as the optimization or evolution mechanism for the kernel covariance matrix. However, the covariance matrix of the CMA-ES is very different in nature to the kernel covariance matrix in online prediction modeling, and therefore the control parameters of the selection and recombination operators in the CMA-ES must be carefully modified to match with the online prediction modeling procedure \cite{hansen2014principled}. The other approach directly uses a CMA-ES algorithm to optimize the kernel covariance matrix based on a prediction performance related objective function, which preserves the well-designed evolution strategy of the CMA-ES \cite{DBLP:journals/corr/abs-1904-07801}. We also use the CMA-ES directly in the optimization of the general symmetric covariance matrix in this paper, and its effects on the prediction performance, in terms of both the sparsification techniques and weight vector updating approaches, are studied.

In order to sufficiently and timely capture the underlying time-varying dynamic characteristics in nonstationary time series, a variety of optimization strategies can be implemented to update both the structure parameters and weight vector. One representative approach to track the nonstationarity is organically combining the construction and elimination procedures of the kernel dictionary selection, which can be achieved by properly setting the sparsification criteria \cite{1356018,6227361}. Another representative approach is intermittently replacing the elements of existing kernel dictionary with new arrived input data, and hence it brings the beneficial property of relative-fixed kernel dictionary size \cite{5495350,7310878,9040439}. For the weight vector adaptation procedure, forgetting mechanisms are typically introduced into the objective function to enhance the tracking ability, such as the sliding-window approach \cite{1661394}, the exponential state forgetting factor \cite{4907077}, and the multi-innovations based approaches \cite{5299173,xu2012robust}. A common observation of the above optimization strategies is that an online sparsification technique is designed for the entire set of all kernel regressors, and the procedure of weight vector updating is also always implemented for the whole set of kernel regressors. In this paper, we propose to divide all the kernel regressors into different groups with multiple kernel connection modes. Different sparsification criteria of kernel dictionary selection can be adopted for the different groups of kernel regressors, and the online weight updating procedures for the weight vectors of different groups can also be different. This generalized optimization strategy with multiple kernel connection modes is highly flexible and effective.

Inspired by the beneficial property of the prediction-error compensation principle and the connectionist representational schemes \cite{finding,sep-connectionism}, we design a generalized optimization strategy to sequentially construct the kernel dictionary in multiple kernel connection modes. This generalized optimization strategy provides a more self-contained way for constructing the entire kernel connections, which enhances the ability to adaptively track the changing dynamic characteristics due to nonstationarity. Three connection modes of kernels are established to organically combine their complementary prediction abilities by using the time-varying prediction-error time series. In summary, the main contribution of this paper is to propose a structure parameter optimized kernel based online prediction approach with a generalized optimization strategy for nonstationary time series, which includes the following specific aspects.
\begin{itemize}
\item [1)] In the light of CMA-ES, the intermittent optimization of the real symmetric form of kernel covariance matrix is realized, which not only improves the kernel structure's flexibility by utilizing the cross relatedness of input variables but also partly alleviates the prediction uncertainty that caused by the kernel dictionary selection for nonstationary time series.
\item [2)] A generalized optimization strategy is designed to construct the kernel dictionary sequentially with multiple kernel connection modes, which provides a more self-contained way for constructing the entire kernel connections and hence enhances the ability to adaptively track the changing dynamic characteristics.
\item [3)] The proper handling of several key parameters is studied in numerical simulations, which has shown the important implications of these parameters for improving the prediction performance of online kernel modeling. 
\end{itemize}

The rest of this paper is organized as follows. We start by surveying representative online kernel algorithms for their kernel dictionary selection procedure and weight vector updating procedure in Section~\ref{sec:2}. In the context of CMA-ES, Section~\ref{sec:3} realizes the intermittent optimization of the real symmetric Gaussian kernel covariance matrix. The generalized optimization strategy with multiple kernel connection modes is presented in Section~\ref{sec:4}. Section~\ref{sec:5} summarizes the proposed online modeling approach and discusses the potential benefits and some unsolved issues. Section~\ref{sec:6} examines the effectiveness of the proposed approach with numerical simulations. The paper concludes with a summary in Section~\ref{sec:7}.

\section{Sparsification Techniques Based Online Kernel Modeling}\label{sec:2}

In this section, two representative online prediction algorithms of sparsification techniques are discussed, in terms of their kernel dictionary selection procedure and weight vector updating procedure. 

\subsection{Procedure of Kernel Dictionary Selection}\label{S2-1}

Assume that we are sequentially given a stream of input-output data pairs $\big\{\bm{x}_n,y_n\big\}_{n=1}^{N}$, with the $p_x$-dimensional input time series $\bm{x}_n\in\mathbb{R}^{p_x \times 1}$ and the corresponding output time series $y_n\in\mathbb{R}$. The input-output data can be denoted as $\bm{X} = \big[\bm{x}_1 ~ \bm{x}_2 \cdots \bm{x}_N\big]^{\rm T}\in \mathbb{R}^{N\times p_x}$ and $\bm{y} = \big[y_1 ~ y_2 \cdots y_N\big]^{\rm T}\in \mathbb{R}^{N \times 1}$. We describe the mapping procedure as
\begin{equation}\label{eq1}
\bm{\varphi} : \quad \mathcal{X} \rightarrow \mathcal{H}, \quad \bm{x} \rightarrow \bm{\varphi}(\bm{x})
\end{equation}
where the high-dimensional or infinite-dimensional Hilbert space $\mathcal{H} = \{ \bm{\varphi}(\bm{x}) \vert \bm{x} \in \bm{X}\}$. A linear combination of the selected kernels at time step $n$ can be obtained as the prediction function
\begin{equation}\label{predictor} 
f(\bm{x}) = \sum_{i=1}^m \widetilde{\alpha}_i k\big(\widetilde{\bm{x}}_i,\bm{x}\big)
\end{equation} 
where $D(n)=\big\{\widetilde{\bm{x}}_i\big\}_{i=1}^m$ denotes the selected kernel dictionary, $m \ll N$ is the size of the selected kernel dictionary, the reproduced kernel function $k$ is defined by as $k\big(\bm{x}_i,\bm{x}_j\big)=\langle\bm{\varphi}(\bm{x}_i),\bm{\varphi}(\bm{x}_j)\rangle$, and the $m$-dimensional weight vector is given by $\widetilde{\bm{\alpha}}=\big[\widetilde{\alpha}_1 ~\widetilde{\alpha}_2\cdots \widetilde{\alpha}_m\big]^{\rm T}$.

\subsubsection{KAF Sparsification Criteria}

The ALD criterion of \cite{1315946} considers the linear dependency between the selected kernels in the form of $\boldsymbol{\varphi}(\mathbf{x})$ and its condition is defined as 
\begin{equation}\label{eq3}
\delta_1(n) = \min \Bigg\| \sum_{i=1}^m \alpha_i(n) \bm{\varphi}\big(\widetilde{\bm{x}}_i\big) - \bm{\varphi}\big(\bm{x}_n\big) \Bigg\|^2 \le \nu_1
\end{equation}  
where $\bm{\alpha}(n)=\big[\alpha_1(n) ~ \alpha_2(n)\cdots \alpha_m(n)\big]^{\rm T}$ is the coefficient vector to form a linear combination of the selected kernels in the form of $\big\{\bm{\varphi}\big(\widetilde{\bm{x}}_i\big)\big\}_{i=1}^m$, and $\nu_1$ denotes the given threshold. Performing the minimization (\ref{eq3}), we can not only check whether this condition is satisfied but also obtain the optimal coefficient vector $\bm{\alpha}(n)$, hence acquiring
\begin{align}
\bm{\alpha}(n) =& \widetilde{\bm{K}}^{-1}(n-1)\widetilde{\bm{k}}_{n-1}\big(\bm{x}_n\big) \label{eq4} \\
\delta_1(n) =& k_{n,n} - \widetilde{\bm{k}}^{\rm T}_{n-1}\big(\bm{x}_n\big) \bm{\alpha}(n) \label{eq5}
\end{align}
where $\widetilde{\bm{K}}(n-1)\in \mathbb{R}^{m\times m}$ is the reduced kernel matrix whose $(i,j)$-th element is derived from the selected kernel dictionary as $\widetilde{K}_{i,j}(n-1)=k\big(\widetilde{\bm{x}}_i,\widetilde{\bm{x}}_j\big)$ for $\widetilde{\bm{x}}_i,\widetilde{\bm{x}}_j \in D(n-1)$, $\widetilde{\bm{k}}_{n-1}\big(\bm{x}_n\big)\in \mathbb{R}^{m\times 1}$ whose $i$-th element is $k\big(\widetilde{\bm{x}}_i,\bm{x}_n\big)$, and $k_{n,n}=k\big(\bm{x}_n,\bm{x}_n\big)$. Consequently, for every $n$, we have
\begin{align}
& \bm{\varphi}\big(\bm{x}_n\big) = \sum_{i=1}^{m(n)} \alpha_i(n) \bm{\varphi}\big(\widetilde{\bm{x}}_i\big) + \bm{\varphi}^{res}(n) \label{eq6} \\
& \big\|\bm{\varphi}^{res}(n)\big\|^2 \le \nu_1 \label{eq7} \\
& \bm{\Phi}_n = \widetilde{\bm{\Phi}}_n \bm{A}^{\rm T}(n) + \bm{\Phi}_n^{res} \label{residual} 
\end{align}
where $\bm{\varphi}^{res}(n)$ denotes the residual component vector, $\bm{\Phi}_n=\big[\bm{\varphi}(\bm{x}_1) ~\bm{\varphi}(\bm{x}_2)\cdots \bm{\varphi}(\bm{x}_n)\big]$, $\widetilde{\bm{\Phi}}_n=\big[\bm{\varphi}\big(\widetilde{\bm{x}}_1\big) ~ \bm{\varphi}\big(\widetilde{\bm{x}}_2\big)\cdots \bm{\varphi}\big(\widetilde{\bm{x}}_m\big)\big]$, and $\bm{\Phi}_n^{res}=\big[\bm{\varphi}^{res}(1) ~ \bm{\varphi}^{res}(2)\cdots \bm{\varphi}^{res}(n)\big]$, while $\bm{A}(n)=\left[\bm{A}^{\rm T}(n-1) ~ \bm{\alpha}(n)\right]^{\rm T}\in \mathbb{R}^{n\times m}$ is the ALD criterion produced coefficients matrix. Note that we have explicitly indicate that $m$ depends on $n$ by using $m(n)$ in (\ref{eq6}).

In \cite{6515200}, the distance criterion of the quantized kernel recursive least squares (QKRLS) algorithm merely depends on the Euclidean distance and does not take into account the distribution of the observed data. This distance criterion is believed to be based on the principle that the principal neighborhoods of data-clustered regions can be approximately represented by the selected kernel dictionary in the Hilbert space $\mathcal{H}$. This criterion is defined as 
\begin{equation}\label{eq9}
\begin{aligned}
\delta_2(n) &= \big\| \bm{x}_n - \widetilde{\bm{x}}_{j^{\star}} \big\|^2 ~ \le \nu_2 \\
j^{\star} &= \arg\min\limits_{1\le j \le m}\big\| \bm{x}_n - \widetilde{\bm{x}}_j \big\|^2 
\end{aligned}
\end{equation} 
where $\nu_2$ denotes the given threshold.

In \cite{FAN20132174}, the changed significance of the loss function can be used to evaluate the significance of each observed input data as a kernel dictionary member. The criterion works in a supervised way and is defined as 
\begin{equation}\label{eq10}
\delta_3(n) = \frac{1}{2}\Delta\widetilde{\bm{\alpha}}^{\rm T} \bm{H}_l(n) \Delta\widetilde{\bm{\alpha}} ~ \le \nu_3
\end{equation}  
where $\Delta\widetilde{\bm{\alpha}}$ is the changed value of the weight vector, $\bm{H}_l(n)$ denotes the Hessian matrix of the loss function, and $\nu_3$ is the given threshold.

\subsubsection{Tunable RBF Sparsification Approaches}

With a fixed kernel dictionary size, the online tunable RBF algorithms of \cite{6334482,7310878,9040439} construct the initial kernel dictionary through two approaches. One approach depends on the distribution of the recent observed input data with the centers based on the nearest-neighbor, or the randomly generated centers. The other approach is the OFS procedure, which adopts the error reduction ratio to evaluate the significance of each training input data as a kernel dictionary member. The OFS procedure works in a supervised way and is described as follows \cite{9040439}.

Given the training data $\{\bm{x}_t,y_t\}_{t=1}^{N_t}$, the full $N_t$-term relationship between the output of the predictor and the actual output $\bm{y}_{N_t}=\big[y_1 ~ y_2\cdots y_{N_t}\big]^{\rm T}$ can be expressed as 
\begin{equation}\label{ofsfunction} 
\bm{y}_{N_t} =  \bm{K}_{N_t} \bm{\alpha}_{N_t} + \bm{e}_{N_t} 
\end{equation}
where $\bm{K}_{N_t}\in \mathbb{R}^{N_t\times N_t}$ is the full regression matrix of the training input data whose $(i,j)$-th element is $k(\bm{x}_i,\bm{x}_j)$, $\bm{\alpha}_{N_t}$ is the full weight vector acquired by the least squares algorithm, and $\bm{e}_{N_t}$ denotes the error vector. 

Using the orthogonal least squares algorithm \cite{CHEN20092670}, the orthogonal decomposition of $\bm{K}_{N_t}$ can be expressed as $\bm{K}_{N_t}= \bm{W}_{N_t}\bm{A}_{N_t}\! =\! \big[\bm{w}_1 ~ \bm{w}_2\cdots \bm{w}_{N_t}\big] \bm{A}_{N_t}$, where $\bm{w}_t$, $1\le t\le N_t$, are the set of orthogonal bases and $\bm{A}_{N_t}$ is an unit upper triangular matrix. Then the expression (\ref{ofsfunction}) can be rewritten as 
\begin{equation}\label{eq12}
\bm{y}_{N_t} =  \bm{W}_{N_t} \bm{g}_{N_t} + \bm{e}_{N_t}   
\end{equation}
where $\bm{g}_{N_t}=\bm{A}_{N_t}\bm{\alpha}_{N_t}=\big[g_1 ~ g_2 \cdots g_{N_t}\big]^{\rm T}$ with $g_i=\bm{w}_i^{\rm T} \bm{y}_{N_t}/\big(\bm{w}_i^{\rm T} \bm{w}_i\big)$ for $1 \le i \le \mathnormal{N}_t$. From (\ref{eq12}),  the sum of squares of the output $\bm{y}_{N_t}$ is given by 
\begin{equation}\label{eq13}
\bm{y}_{N_t}^{\rm T} \bm{y}_{N_t} = \sum_{j=1}^{N_t} g_j^2 \bm{w}_j^{\rm T}\bm{w}_j + \bm{e}_{N_t}^{\rm T} \bm{e}_{N_t}
\end{equation}
Therefore, the error reduction ratio to evaluate the significance of each training input data can be defined as 
\begin{equation}\label{eq14}
[\text{err}]_j = \frac{g_j^2 \bm{w}_j^{\rm T} \bm{w}_j}{\bm{y}_{N_t}^{\rm T} \bm{y}_{N_t}} 
\end{equation}

Due to the unit upper triangular form of $\bm{A}_{N_t}$ and the sequential computation of $\bm{w}_j$ in the orthogonal least squares algorithm \cite{orthogonal}, the whole OFS procedure also can be performed in a recursive way as in the ALD kernel recursive least squares (ALD-KRLS) algorithm, supported with the kernel dictionary updating approach \cite{5495350}. The other approach that depends on the distribution of recent observed input data also can be performed in a recursive way. Therefore, the recursive operating ways of kernel dictionary selection are unified in the two representative types of online prediction algorithms.   

\subsubsection{Analysis of Kernel Dictionary Selection Procedure}

The kernel dictionary selection is indeed an objective-oriented problem, which leads to the diverse sparsification criteria and the organic combination of different sparsification approaches \cite{7120172,8291833,9147041}. According to the representer theorem \cite{sch2001a}, the optimal solution to the minimization of the regularized empirical risk can be expressed as weighted kernels that composed of all the available training samples. Then the sparsification can be viewed as a procedure to deal with the differences between the two prediction spaces, one is formed by all the available training samples and the other is formed by the selected kernel dictionary. Considering that the high-dimensional or infinite-dimensional Hilbert space $\mathcal{H}$ may cause the excessive growth of kernel dictionary, diverse sparsification approaches also can help to properly construct the kernel dictionary in order to obtain better generalization ability. From this perspective, sparsification in the kernel dictionary selection procedure becomes an effective measure of parameter regularization for superior prediction performance.

Once a sparsification approach is chosen, it is vital to properly set the threshold of the sparsification criterion and the termination condition of the kernel dictionary selection. The size of kernel dictionary is determined by the termination condition, which plays a crucial role in controlling the model generalization ability. As the sequential data arrive, the final selected kernel dictionary depends on both the first selected member $\widetilde{\bm{x}}_1$ and the given threshold for the sparsification criterion, if other parameters in the kernel dictionary selection procedure have already been properly set. For the methods of deciding the thresholds, the reader is referred to the aforementioned corresponding references. It is worth recapping that the first selected member $\widetilde{\bm{x}}_1$ can make much difference to the final selected kernel dictionary, since the selected kernel dictionary is generated in a recursive way and all the subsequent selected members should fulfill the criteria that established by the already existed kernel dictionary. This is especially true for the nonstationary time series. 

\subsection{Procedure of Weight Vector Updating}\label{S2-2}

We now discuss the weight vector updating procedures in the representative online prediction algorithms. The weight vector updating approaches of online kernel modeling can be classified into two categories, based on whether or not to explicitly consider the property of the mapping function $\bm{\varphi}(\bm{x})$ in (\ref{predictor}). Most KAF algorithms \cite{1315946,6515200,FAN20132174} consider the property of $\bm{\varphi}(\bf{x})$ in the weight vector updating procedure and the solution of $\widetilde{\bm{\alpha}}$ can be acquired by minimizing the corresponding loss function. For the online tunable RBF algorithms \cite{7310878,9040439} and the robust online kernel learning algorithm \cite{7407373}, the solution of $\widetilde{\bm{\alpha}}$ is acquired with the information matrix of the kernel vector, and they do not explicitly consider the property of the mapping function $\bm{\varphi}(\bm{x})$.

\subsubsection{Weight Vector Updating in KAFs}

Considering for example the ALD-KRLS algorithm, the sparsification procedure at time step $n$ can be denoted as 
\begin{equation}\label{eq15}
f(\bm{x}) = \sum_{i=1}^n \alpha_i(n) k\big(\bm{x}_i,\bm{x}\big) \rightarrow \sum_{i=1}^m \widetilde{\alpha}_i(n) k\big(\widetilde{\bm{x}}_i,\bm{x}\big)
\end{equation}
With the selected kernel dictionary, the loss function $\mathcal{L}(\bm{\omega}_{n})$ can be defined as
\begin{equation}\label{eq16}
\mathcal{L}(\bm{\omega}_n) = \Big\|\bm{y}_n -\bm{\Phi}_n \bm{\omega}_n\Big\|^2 
\end{equation}
with $\bm{y}_n=\big[y_1 ~ y_2 \cdots y_n\big]^{\rm T}$. The optimal solution of $\bm{\omega}_n$ that minimizes $\mathcal{L}(\bm{\omega}_{n})$ can be expressed as 
\begin{equation}\label{optimal} 
\bm{\omega}_n = \sum_{i=1}^n \alpha_i(n) \bm{\varphi}\big(\bm{x}_i\big) = \bm{\Phi}_n \bm{\alpha}(n)
\end{equation}
By omitting the residual component vector $\boldsymbol{\varphi}^{res}(n)$ in (\ref{residual}), we have the approximation $\bm{\omega}_n=\bm{\Phi}_n\bm{\alpha}(n) \approx \widetilde{\bm{\Phi}}_n\bm{A}^{\rm T}(n) \bm{\alpha}(n)=\widetilde{\bm{\Phi}}_n\widetilde{\bm{\alpha}}(n)$, where $\widetilde{\bm{\alpha}}(n)=\bm{A}^{\rm T}(n)\bm{\alpha}(n)$. Then the loss function $\mathcal{L}\big(\widetilde{\bm{\alpha}}(n)\big)$ can be defined as
\begin{align}\label{eq18}
\mathcal{L}\big(\widetilde{\bm{\alpha}}(n)\big) =& \Big\|\bm{y}_n - \bm{\Phi}_n^{\rm T}\widetilde{\bm{\Phi}}_n\widetilde{\bm{\alpha}}(n)\Big\|^2 \nonumber \\
=& \Big\|\bm{y}_n -\bm{A}(n)\widetilde{\bm{K}}(n)\widetilde{\bm{\alpha}}(n)\Big\|^2 
\end{align}
The optimal weight vector $\widetilde{\bm{\alpha}}(n)$ can be directly obtained by the least squares algorithm as
\begin{align}\label{eq19}
\widetilde{\bm{\alpha}}(n) =& \left(\bm{A}(n)\widetilde{\bm{K}}(n)\right)^{\dag}\bm{y}_n \notag \\ 
=& \widetilde{\bm{K}}(n)^{-1}\left(\bm{A}^{\rm T}(n)\bm{A}(n)\right)^{-1}\bm{A}^{\rm T}(n)\bm{y}_n  
\end{align}

In \cite{1315946}, the ALD-KRLS algorithm realizes the recursive form of updating the weight vector $\widetilde{\bm{\alpha}}(n)$ in both the cases that the kernel dictionary $D(n)$ is expanded and is not expanded.

\subsubsection{Weight Vector Updating in Tunable RBF Algorithms}

In \cite{7310878,9040439}, the weight vector $\widetilde{\bm{\alpha}}$ in (\ref{predictor}) is updated with the multi-innovation recursive least squares (MRLS) algorithm, which uses the latest $p$ innovations to form the regression information matrix. The cumulative loss function $\mathcal{L}\big(\widetilde{\bm{\alpha}}(n)\big)$ in this case is defined as
\begin{equation}\label{cumulative} 
\mathcal{L}\big(\widetilde{\bm{\alpha}}(n)\big) = \frac{1}{2}\sum_{i=n-p+1}^n \beta^{n-i}\left(y_i - \widetilde{\bm{k}}_n^{\rm T}(\bm{x}_i) \widetilde{\bm{\alpha}}(n)\right)^2
\end{equation}
where $\beta$ is the forgetting factor. The associated information matrix is $\widetilde{\bm{K}}_p=\big[\widetilde{\bm{k}}_n(\bm{x}_{n-p+1}) ~ \widetilde{\bm{k}}_n(\bm{x}_{n-p+2})\cdots \widetilde{\bm{k}}_n(\bm{x}_n)\big]^{\rm T}$, and the optimal weight vector $\widetilde{\bm{\alpha}}(n)$ can be directly obtained by the recursive least squares algorithm:
\begin{align}\label{e21}
\left\{\begin{array}{ccl}
\bm{\Psi}_n & = & \bm{P}_{n-1} \widetilde{\bm{K}}_p^{\rm T} \big(\beta \bm{I}_p + \widetilde{\bm{K}}_p \bm{P}_{n-1} \widetilde{\bm{K}}_p^{\rm T}\big)^{-1} \\
\bm{P}_n & = & \big(\bm{P}_{n-1} - \bm{\Psi}_n \widetilde{\bm{K}}_p \bm{P}_{n-1}\big) \beta^{-1} \\
\widetilde{\bm{\alpha}}(n) & = & \widetilde{\bm{\alpha}}(n-1) + \bm{\Psi}_n \bm{e}_p
\end{array}\right.
\end{align}
where $\bm{\Psi}_n\in \mathbb{R}^{m \times p}$ is the Kalman gain matrix, $\bm{P}_n \in \mathbb{R}^{m\times m}$ is the inverse of the covariance matrix updated by the information matrix $\widetilde{\bm{K}}_p$, $\bm{I}_p$ denotes the $p\times p$ identity matrix, and $\bm{e}_p$ is the error vector of the latest $p$ predictors.

\subsubsection{Weight Vector Updating in Robust Online Kernel Learning}

Instead of using the cumulative loss function $\mathcal{L}\big(\widetilde{\bm{\alpha}}(n)\big)$ (\ref{cumulative}), the robust recurrent kernel online learning (RRKOL) \cite{7407373} utilizes the instantaneous prediction error $e_n^2$, together with the extra information provided by the past recurrent feedback signals $y_{n-1},y_{n-2},\cdots,y_{n-d}$ that are included in the input variable $\bm{x}_n$. According to the chain rule, the full derivative of the instantaneous prediction error $e_n^2$ is used to iteratively update the weight vector $\widetilde{\bm{\alpha}}(n)$
\begin{equation}\label{derivative} 
\left.\frac{\mathrm{d} e_n^2}{\mathrm{d} \widetilde{\bm{\alpha}}^{\rm T}(n)}\right|_{\bm{\Lambda}_n} = \frac{\partial \big(e_n^2\big)}{\partial \widetilde{\bm{\alpha}}^{\rm T}(n)} + \frac{\partial \big(e_n^2\big)}{\partial \widetilde{\bm{k}}_n} \frac{\partial \widetilde{\bm{k}}_n}{\partial \widetilde{\bm{\alpha}}^{\rm T}(n)} \bm{\Lambda}_n
\end{equation}
where $\bm{\Lambda}_n$ is the hyperparameter matrix to weight the second recurrent term. The distinct attribute of the RRKOL algorithm is the second recurrent term in (\ref{derivative}) which considers the past recurrent feedback signals within $\bm{x}_n$. It provides a distinct insight into what role the past recurrent feedback signals can play in the weight vector updating procedure.

Similarly, the weight vector updating is also an objective-oriented problem. Whether or not to consider the property of the mapping function $\bm{\varphi}(\bm{x})$ may bring differences to the handling of (\ref{optimal}) and further lead into some uncertainty about prediction performance. It is not advisable to aimlessly use the property of the mapping function $\boldsymbol{\varphi}(\mathbf{x})$ if the awareness of how the property impacts on the prediction performance is insufficient. Moreover, other weight vector updating techniques, such as the forgetting mechanisms, the regularized $\mathcal{L}\big(\widetilde{\bm{\alpha}}\big)$ and the sliding window/multi-innovation, also can be properly chosen to be implemented for specific online modeling targets. In order to sufficiently capture the underlying dynamic characteristics in nonstationary time series, it is necessary to carefully design an optimization strategy to achieve the effective combination of these weight vector updating techniques.

\section{Intermittent Optimization of Kernel Covariance Matrix in the light of CMA-ES}\label{sec:3}

Assume that the type of kernel is chosen a priori, and without loss of generality we take the Gaussian kernel as an example. With the aid of the well-studied Gaussian kernel distribution, the important role of the kernel covariance matrix is clearly understood, in terms of shaping the prediction model \cite{bishop2006pattern, DBLP:journals/corr/Hansen16a}. The Gaussian kernel based prediction function (\ref{predictor}) can be described as
\begin{align}\label{Gpredictor} 
f(\bm{x}) =& \sum_{i=1}^m \widetilde{\alpha}_i k\big(\widetilde{\bm{x}}_i,\bm{x}\big) \notag \\
=& \sum_{i=1}^m \widetilde{\alpha}_i \exp\left( \big(\bm{x} - \widetilde{\bm{x}}_i\big)^{\rm T} \bm{\Sigma}_i^{-1}\big(\bm{x} - \widetilde{\bm{x}}_i\big) / h_0\right) 
\end{align}
where $\bm{\Sigma}_i\in \mathbb{R}^{p_x \times p_x}$ is the kernel covariance matrix of the $i$-th Gaussian kernel member and $h_0$ is an order of magnitude factor. The quadratic term $\Delta^2_i$ can be expressed as
\begin{align}\label{expansion} 
\Delta^2_i =& \big(\bm{x} - \widetilde{\bm{x}}_i\big)^{\rm T} \bm{\Sigma}_i^{-1}\big(\bm{x} - \widetilde{\bm{x}}_i\big) \notag \\
=& \Big(\big(\bm{x} - \widetilde{\bm{x}}_i\big)^{\rm T} \bm{U}_i\Big)\Big(\bm{U}_i^{\rm T} \big(\bm{x} - \widetilde{\bm{x}}_i\big)\Big)
\end{align}
where $\bm{U}_i=\big[ \bm{d}_{i,1}/\sqrt{\lambda_{i,1}} ~ \bm{d}_{i,2}/\sqrt{\lambda_{i,2}}\cdots \bm{d}_{i,p_x}/\sqrt{\lambda_{i,p_x}}\big]$, with $\lambda_{i,j}$ and $\bm{d}_{i,j}$, $1\le j\le p_x$, denoting the eigenvalues and the corresponding eigenvectors of $\bm{\Sigma}_i$, respectively, and the eigenvectors can be chosen to form an orthonormal set. The matrix $\bm{U}_i$ can be interpreted as a coordinate transformation of shifting and rotating with respect to the original coordinates \cite{bishop2006pattern}. The coordinate transformation directly shapes the surface of the prediction function (\ref{Gpredictor}) and further influences the prediction performance. 

The following weighted selection mechanism estimates the empirical kernel covariance matrix $\bm{\Sigma}^{emp}$ using selected observed sequential data set $\big\{\widehat{\bm{x}}_j\big\}_{j=1}^{N_e}$
\begin{equation}\label{empirical} 
\bm{\Sigma}_i^{emp} = h_0 \sum_{j=1}^{N_e} \widehat{\omega}_j \big(\widehat{\bm{x}}_j - \widetilde{\bm{x}}_i\big)\big(\widehat{\bm{x}}_j - \widetilde{\bm{x}}_i\big)^{\rm T}
\end{equation}
where $\widehat{\bm{\omega}}=\big[\widehat{\omega}_1 ~ \widehat{\omega}_2 \cdots \widehat{\omega}_{N_e}\big]^{\rm T}$ denotes the weight vector. The sequential data sets can be selected based on the distribution of the observed samples, such as the widely-adopted nearest neighbor method for RBF kernels \cite{7310878,9040439,KITAYAMA20114726}. This class of estimation approaches use the selected sequential data sets to approximately represent the intended population of each Gaussian kernel regressor in (\ref{Gpredictor}), which contributes limited predictive attributes due to its unsupervised way.

To realize the intermittent optimization of the kernel covariance matrix in the context of CMA-ES, the correlations of the kernel covariance matrix with other parameters need to be considered, in terms of both the kernel dictionary selection and weight vector updating. For the kernel dictionary selection procedures discussed in Section~\ref{sec:2}, the actual selected kernel dictionary can be very different from the suboptimal or optimal kernel dictionary, due to the uncertainty caused by the nonstationarity. Given the optimal or suboptimal sparsified kernel dictionary $\big\{\widetilde{\bm{x}}_i^{\star}\big\}_{i=1}^m$, with the corresponding estimated kernel covariance matrices $\bm{\Sigma}_i^{\star}$ and updated weights $\widetilde{\alpha}_i^{\star}$, the prediction function (\ref{Gpredictor}) becomes
\begin{align}\label{optipredictor} 
f^{\star}(\bm{x}) =& \sum_{i=1}^m \widetilde{\alpha}_i^{\star} \exp\left( \big(\bm{x} - \widetilde{\bm{x}}_i^{\star}\big)^{\rm T} {\bm{\Sigma}_i^{\star}}^{-1}\big(\bm{x} - \widetilde{\bm{x}}_i^{\star}\big) / h_0\right) \notag \\
=& \sum_{i=1}^m \widetilde{\alpha}_i^{\star} \exp\left( \big(\bm{x} - \widetilde{\bm{x}}_i\big)^{\rm T} \bm{C} {\bm{\Sigma}_i^{\star}}^{-1} \bm{C}^{\rm T} \big(\bm{x} - \widetilde{\bm{x}}_i\big) / h_0\right) \notag \\
=& \sum_{i=1}^m \widetilde{\alpha}_i^{\star} \exp\left( \big(\bm{x} - \widetilde{\bm{x}}_i\big)^{\rm T} \widetilde{\bm{\Sigma}}_i^{-1}\big(\bm{x} - \widetilde{\bm{x}}_i\big) / h_0\right) 
\end{align}
where $\bm{C}^{\rm T}$ is the transformation matrix from $\big(\bm{x} - \widetilde{\bm{x}}_i\big)$ to $\big(\bm{x} - \widetilde{\bm{x}}_i^{\star}\big)$. Comparing the prediction functions in (\ref{Gpredictor}) and (\ref{optipredictor}), it can be seen that the object to be optimized should be $\widetilde{\bm{\Sigma}}_i^{-1}=\bm{C}{\bm{\Sigma}_i^{\star}}^{-1}\bm{C}^{\rm T}$, not $\big(\bm{\Sigma}_i^{emp}\big)^{-1}$ of (\ref{empirical}), if the kernel dictionary selection procedure and the weight vector updating procedure are taken into consideration. The optimization of the real symmetric matrix $\widetilde{\bm{\Sigma}}_i^{-1}$ greatly improves the kernel structure's flexibility than the restricted form of $\bm{\Sigma}_i^{-1}$. Similar to the principal components in principal component analysis (PCA) method or the orthogonal search paths in evolutionary algorithms \cite{6790375}, the enhanced kernel structure flexibility enhances the ability of each kernel regressor in (\ref{optipredictor}) to capture the underlying dynamic characteristics in the neighborhood.

For notational simplification, we drop the subindex $i$ in the sequel. The real symmetric matrix $\widetilde{\bm{\Sigma}}^{-1}$ can be optimized in the following ``rank-one updating'' form
\begin{equation}\label{optiform} 
\widetilde{\bm{\Sigma}}^{-1}(n) = (1 - c_0)\widetilde{\bm{\Sigma}}^{-1}(n-1) \pm \bm{p}_{\sigma}(n)\bm{p}_{\sigma}^{\rm T}(n) 
\end{equation}
where $c_0$ is the learning rate which can be calculated as $c_0 = 2/p_x^2$ \cite{DBLP:journals/corr/Hansen16a}, and $\bm{p}_{\sigma}$ is the target vector to optimize the kernel covariance matrix. In order to give more predictive attributes to the intermittent optimization of $\widetilde{\bm{\Sigma}}^{-1}$, the objective function to be minimized is set to
\begin{equation}\label{objfunc} 
\mathcal{L}_{\sigma} = \sum_{\bm{x}_j \in D_{\sigma}} \omega_{\sigma_j} \big(y_j - f^{\star}(\bm{x}_j)\big)^2 
\end{equation} 
where $D_{\sigma}$ is the set of the selected samples, and $\omega_{\sigma_j}$ are the weights of the loss to tradeoff the local and global characteristics. To better prevent the occurrence of some irregularities, such as outliers, during the online prediction procedure, the selected samples $D_{\sigma}$ in (\ref{objfunc}) and the order of magnitude factor $h_0$ in (\ref{optipredictor}) can be used to deal with these irregularities.

According to the above analysis, we focus on directly using the CMA-ES to optimize the target vector $\bm{p}_{\sigma}$ in (\ref{optiform}) based on the objective function $\mathcal{L}_{\sigma}$ of (\ref{objfunc}). We introduce the so-called ``pure CMA-ES'' algorithm to better describe the intermittent optimization of $\bm{p}_{\sigma}$. This pure CMA-ES algorithm is the fundamental and simplest version among many variants of the CMA-ES algorithms. As an evolutionary algorithm, it consists of three steps, the initialization step, the repeating evolution step with four operators of mutation, evaluation, selection and recombination, and the termination step \cite{bck2013contemporary}. 

Let the population size be $\lambda_c$. A population of new individuals $\bm{x}_k^{(g+1)}$ at the $(g+1)$-th generation are generated by the mutation operator, which samples a multivariate normal distribution $\mathcal{N}\big(\bm{0},\bm{C}^{(g)}\big)$ with the zero mean vector $\bm{0}$ and the covariance matrix $\bm{C}^{(g)}$. The sampling operation, from the $g$-th generation to the next generation, can be described as
\begin{equation} \label{sampling} 
\bm{x}_k^{(g+1)} \sim \bm{m}^{(g)} + \sigma_c^{(g)}\mathcal{N}\big(\bm{0},\bm{C}^{(g)}\big)
\end{equation} 
where $\bm{m}^{(g)}$ denotes the mean vector of the $g$-th population individuals, and $\sigma_c^{(g)}$ is the mutation step-size. Then, evaluation of these mutated individuals on the objective function $\mathcal{L}_{\sigma}$ (\ref{objfunc}) is implemented,  and the top $\mu_c$ ranked individuals, $\bm{x}_{k(1)}^{(g+1)}, \bm{x}_{k(2)}^{(g+1)},\cdots, \bm{x}_{k(\mu_{c})}^{(g+1)}$, are selected, where $\mu_c < \lambda_{c}$. Finally, the weighted recombination of the best $\mu_{c}$ ranked individuals is reflected in the following updating equations for the  parameters of the mutation operator (\ref{sampling})
\begin{align}
\bm{m}^{(g+1)} =& \bm{m}^{(g)} + c_m\sum_{i=1}^{\mu_c} \omega_m(i)\Big(\bm{x}_{k(i)}^{(g+1)} - \bm{m}^{(g)}\Big)  \label{eq30} \\
\bm{C}^{(g+1)} =& \Bigg(\!\! 1 - c_1 - c_{\mu} \sum_{i=1}^{\lambda_c} \omega_c(i)\!\! \Bigg) \bm{C}^{(g)}\! + c_1 \underbrace{\bm{p}_{c1}^{(g+1)}\bigl(\bm{p}_{c1}^{(g+1)}\bigr)^{\rm T}}_\text{rank-one update} \notag \\ 
& + c_{\mu}\underbrace{\sum_{i=1}^{\lambda_c} \omega_c(i) \bm{p}_{c\mu(i)}^{(g+1)} \big(\bm{p}_{c\mu(i)}^{(g+1)}\big)^{\rm T}}_\text{rank-$\mu_c$ update} \label{eq31} \\
\sigma_{c}^{(g+1)} =& \sigma_{c}^{(g)}\exp\left(\frac{c_{\sigma_c}}{d_{\sigma_c}}\left(\frac{\| {\bm{p}_{\sigma_c}}^{(g+1)} \|}{\widehat{\chi_{p_x}}} - 1 \right)\right) \label{updationequ} 
\end{align}
where $\bm{p}_{c1}$ and $\big\{\bm{p}_{c\mu(1)}, \bm{p}_{c\mu(2)}, \cdots , \bm{p}_{c\mu(\lambda_{c})}\big\}$ are respectively the well designed evolution paths of the ``rank-one update'' and ``rank-$\mu_{c}$ update'', $\bm{p}_{\sigma_c}$ is the cumulative step-size control evolution path for the mutation step-size, $\bm{\omega}_c=\big[\omega_c(1) \cdots \omega_c(\lambda_c)\big]^{\rm T}$ and $\bm{\omega}_m=\big[\omega_m(1) \cdots \omega_m(\lambda_c)\big]^{\rm T}$ are the respective coefficient vectors, while $c_m$, $c_1$, $c_{\mu}$ and $\frac{c_{\sigma_c}}{d_{\sigma_c}}$ are the learning rates of the respective evolution paths, and $\widehat{\chi_{p_x}}$ is the expected length of a random variable distributed according to $\mathcal{N}\big(\bm{0},\bm{I}_{p_x}\big)$ which helps to normalize the length of $\bm{p}_{\sigma_c}$. The detailed parameter settings can be found in \cite{DBLP:journals/corr/Hansen16a}.

\begin{algorithm}[!t]
\caption{Intermittent Optimization of Kernel Covariance Matrix Using CMA-ES for ALD-KRLS} 
\label{aldcma} 
\begin{algorithmic}
\STATE \textbf{Initialization}
\STATE \quad Initialize strategy parameters of CMA-ES. 
\STATE \quad Initialize $\widetilde{\bm{\Sigma}}^{-1}(0)$ and $c_0$ in (\ref{optiform}).
\STATE \textbf{Repeat} 
\STATE \quad Mutation as in (\ref{sampling}).
\STATE \quad Evaluation
\STATE \quad \quad \textbf{For} each mutation individual
\STATE \quad \quad \quad Set each mutation individual as $\bm{p}_{\sigma}$. 
\STATE \quad \quad \quad Calculate $\widetilde{\bm{\Sigma}}^{-1}$ as (\ref{optiform}). 
\STATE \quad \quad \quad Initialize $h_0$ in (\ref{optipredictor}).
\STATE \quad \quad \quad Initialize parameters as in ALD-KRLS. 
\STATE \quad \quad \quad \textbf{For} each selected sample in $D_{\sigma}$
\STATE \quad \quad \quad \quad Operate kernel dictionary selection procedure.
\STATE \quad \quad \quad \quad Operate weight vector updating procedure.
\STATE \quad \quad \quad \quad Calculate prediction performance.
\STATE \quad \quad \quad \textbf{End for}
\STATE \quad \quad \quad Return value of objective function $\mathcal{L}_{\sigma}$.
\STATE \quad \quad \textbf{End for}
\STATE \quad Selection
\STATE \quad \quad Return best ranked individuals $\bm{x}_{k(1)}^{(g+1)},\cdots,\bm{x}_{k(\mu_c)}^{(g+1)}$.
\STATE \quad Recombination as (\ref{eq30}) to (\ref{updationequ}).
\STATE \textbf{Until} termination criterion is fulfilled
\STATE Calculate $\widetilde{\bm{\Sigma}}^{-1}$ as (\ref{optiform}) with $\bm{p}_{\sigma}=\bm{x}_{k(1)}^{(g+1)}$.
\STATE Return final $\widetilde{\bm{\Sigma}}^{-1}$.
\STATE \textbf{End}
\end{algorithmic}
\end{algorithm}
 
As a randomized search algorithm, the fundamental design principles of the CMA-ES are the invariance, which enables an identical behavior on a class of objective functions, and the unbiasedness, which partly prevents the risk of divergence or premature convergence \cite{DBLP:journals/corr/Hansen16a}. These properties provide algorithmic setting options to be chosen for online modeling in practical applications. The termination criterion in the CMA-ES should be carefully set according to specific online prediction problems. Our algorithm implementations of using the CMA-ES to optimize the kernel covariance matrix $\widetilde{\bm{\Sigma}}^{-1}$ (\ref{optiform}) can be classified into two cases, based on whether the kernel dictionary selective criterion is independent of the calculation of the updated kernel covariance matrix $\widetilde{\bm{\Sigma}}^{-1}$ in (\ref{optipredictor}). As summarized in Algorithm~\ref{aldcma}, the ALD criterion represents the case that the kernel dictionary selective criteria in the online prediction algorithms are dependent on the updated kernel covariance matrix. For Algorithm \ref{discma}, the distance criterion represents the case that the kernel dictionary selective criteria are independent of the updated kernel covariance matrix. 

\begin{algorithm}[!t]
\caption{Intermittent Optimization of Kernel Covariance Matrix Using CMA-ES for QKRLS} 
\label{discma} 
\begin{algorithmic}
\STATE \textbf{Initialization} 
\STATE \quad Initialize parameters as in QKRLS.   
\STATE \quad Initialize $\widetilde{\bm{\Sigma}}^{-1}(0)$ and $c_0$ in (\ref{optiform}).
\STATE \quad Initialize $h_0$ in (\ref{optipredictor}).
\STATE \quad Operate kernel dictionary selection procedure.
\STATE \quad Initialize strategy parameters as in CMA-ES.
\STATE \textbf{Repeat} 
\STATE \quad Mutation as in (\ref{sampling}).
\STATE \quad Evaluation
\STATE \quad \quad \textbf{For} each mutation individual
\STATE \quad \quad \quad Set each mutation individual as $\bm{p}_{\sigma}$. 
\STATE \quad \quad \quad Calculate $\widetilde{\bm{\Sigma}}^{-1}$ as (\ref{optiform}).    
\STATE \quad \quad \quad \textbf{For} each selected sample in $D_{\sigma}$
\STATE \quad \quad \quad \quad Operate weight vector updating procedure.
\STATE \quad \quad \quad \quad Calculate prediction performance.
\STATE \quad \quad \quad \textbf{End for}
\STATE \quad \quad \quad Return value of objective function $\mathcal{L}_{\sigma}$.
\STATE \quad \quad \textbf{End for}
\STATE \quad Selection
\STATE \quad \quad Return best ranked individuals $\bm{x}_{k(1)}^{(g+1)},\cdots,\bm{x}_{k(\mu_c)}^{(g+1)}$. 
\STATE \quad Recombination as (\ref{eq30}) to (\ref{updationequ}).
\STATE \textbf{Until} termination criterion is fulfilled
\STATE Calculate $\widetilde{\bm{\Sigma}}^{-1}$ as in (\ref{optiform}) with $\bm{p}_{\sigma}=\bm{x}_{k(1)}^{(g+1)}$.
\STATE Return final $\widetilde{\bm{\Sigma}}^{-1}$.
\STATE \textbf{End}
\end{algorithmic}
\end{algorithm}

\section{Generalized Optimization Strategy in Kernel Connection Mode}\label{sec:4}

Generally, the prediction-error time series is a good metric, providing a very useful clue to improve the prediction performance of online sequential data \cite{finding}. In the kernel dictionary selection procedure, it is well known that the kernel dictionary size $m$ is of vital importance to the generalization ability of the prediction function (\ref{predictor}), and there usually exists a proper kernel dictionary size for specific observed sequential data. It implies that any part of the whole kernel regressors can be viewed as an error compensator for the other part of the kernel regressors, i.e., the compensator can help to further capture the underlying dynamics in the prediction-error time series that generated by the already existing part. In this paper, we simply call this the principle of prediction-error compensation. Essentially, the prediction-error compensation principle is a product of the relationship between the whole and the parts of the kernel regressors.

\begin{figure}[tb!]
\begin{center}
\vspace*{-1mm}
\includegraphics[width=\columnwidth]{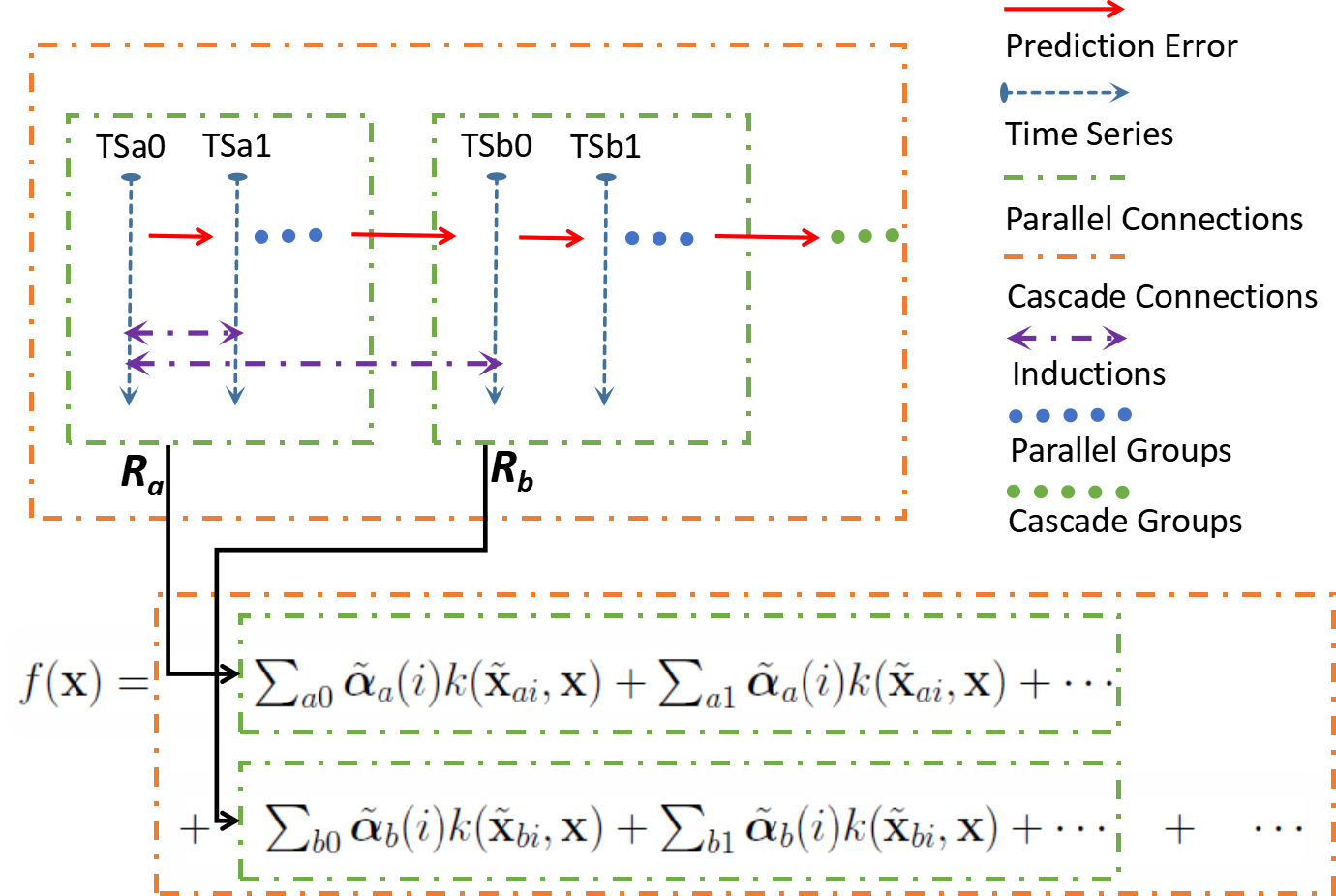}
\end{center}
\vspace*{-3mm}
\caption{Illustration of the generalized connection modes.}
\label{Connections} 
\vspace*{-4mm}
\end{figure}

In order to better realize the prediction-error compensation principle in the kernel dictionary selection procedure, a generalized optimization strategy is designed to sequentially construct the kernel dictionary in multiple kernel connection modes. Inspired by the connectionist representational schemes~\cite{sep-connectionism}, three basic connection modes of kernel regressors are established to organically combine their complementary prediction abilities by using the time-varying prediction-error time series. Similar to the connections in the circuit topology \cite{fundamentals}, two or more kernel regressors are in \emph{series connections} if their kernel dictionary elements are selected from the same time series (input time series or prediction-error time series) and the weights of these kernel regressors form a weight vector to be updated synchronously. Two or more groups of series-connected kernel regressors are in \emph{parallel connections} if their kernel dictionary members are selected from the same time series, and in an order, the weight vector of the next parallel-connected group is to be updated according to the prediction-error time series that generated by the previous parallel-connected group. A cascade-connected group is a head-to-tail arrangement of two or more parallel-connected groups. In \emph{cascade connections}, the kernel dictionary members of the next cascade-connected group are selected from the prediction-error time series that generated by the previous cascade-connected group. As illustrated in the connection diagram of Fig.~\ref{Connections}, the kernel dictionaries of the cascade-connected groups $R_a$ and $R_b$ are selected from the time series $TS_{a0}$ and $TS_{b0}$, respectively. The three basic connection modes can be distinguished in terms of how to realize the prediction-error compensation principle in the kernel dictionary selection procedure.

After the initial construction of the three basic connection modes, specific relationships between different time series can be set, and some of the selected kernel dictionary members also can be transferred into other groups based on the adopted information criteria. In the three basic connection modes, some connections can be left out if not needed. All the adopted optimization strategies in the aforementioned kernel online algorithms can reconstruct their corresponding models, in terms of the three basic connection modes, which motivates us to call the framework of three basic connection modes a generalized optimization strategy.

For the online prediction of nonstationary time series, the generalized optimization strategy provides a more self-contained way to construct the entire kernel connections and thus better explores the complementary prediction abilities in the time-varying prediction-error time series, which enhances the ability to track the changing dynamic characteristics due to nonstationarity. The three basic connection modes divide the whole kernel regressors into different groups, which actually provides a perspective to handle with the relationship of the whole and the parts of kernel regressors. In the procedure of the generalized optimization strategy, the modeling of the next specific connection group can make a change accordingly if the previous connection groups have already provided useful information. More useful information can be acquired by monitoring the prediction performance of all the divided groups than by just monitoring the whole kernel regressors.

\begin{algorithm}[!t]
\caption{Realization of Proposed Generalized Optimization Strategy for Improved Information Interactions} 
\label{completeness} 
\begin{algorithmic}
\STATE Initialization 
\STATE \textbf{Training Procedure}
\STATE Form time series $TS_{a0}$.
\STATE Operate kernel dictionary selection procedure.
\STATE Determine parallel connections of selected kernel dictionary.
\STATE Operate weight vector updating procedure of first parallel-connected group.
\STATE Calculate its prediction performance.
\STATE \textbf{Repeat}
\STATE \quad Form prediction-error time series of previous parallel-
\STATE \quad connected groups.
\STATE \quad Add a new parallel-connected group.
\STATE \quad Operate weight vector updating procedure.
\STATE \quad Calculate its prediction performance.
\STATE \textbf{Until} termination criterion is fulfilled
\STATE Return generated parallel-connected groups as first cascade-connected group.
\STATE \textbf{Repeat} 
\STATE \quad Form prediction-error time series of previous cascade-
\STATE \quad connected groups. 
\STATE \quad Construct kernel dictionary of new cascade group.
\STATE \quad Determine parallel connections of new cascade group.
\STATE \quad Operate weight vector updating procedure of each newly-
\STATE \quad added parallel-connected group.
\STATE \quad Calculate prediction performance. 
\STATE \textbf{Until} termination criterion is fulfilled
\STATE Record generated prediction-error time series.
\STATE Return whole prediction function and its connections.
\STATE Optimize kernel covariance matrices if necessary.
\STATE \textbf{Online Prediction Procedure}
\STATE \textbf{For} each new arriving sample
\STATE \quad Calculate prediction performance of all connections.
\STATE \quad Record all generated prediction-error time series. 
\STATE \quad Operate weight vector updating procedure for each
\STATE \quad parallel-connected group.
\STATE \quad Adjust corresponding kernel connections if necessary.
\STATE \quad Optimize kernel covariance matrices if necessary.
\STATE \quad Record all updated prediction-error time series.
\STATE \quad Return updated prediction function and its connections.
\STATE \textbf{End for}
\end{algorithmic}
\end{algorithm}

\section{Summary of Proposed Approach}\label{sec:5}

Both the intermittent optimization of the kernel covariance matrix and the generalized optimization strategy in three basic kernel connection modes can improve the information interactions between the network topology of kernel regressors and the optimization of model parameters, in both the kernel dictionary selection procedure and the weight vector updating procedure. The improved information interaction not only enhances the ability to deal with various online modeling problems but also provides a more flexible kernel structure and a more compatible way of kernel connections that can be further combined with other existing online modeling techniques. An implementation or realization of the proposed generalized optimization strategy is given in Algorithm~\ref{completeness}. To be more specific, given a nonstationary time series $TS_{a0}$ to be predicted, the operations of Algorithm~\ref{completeness} are as follows.

First the kernel dictionary selection procedure in the first cascade-connected group $R_a$ can be naturally carried out by representative online prediction algorithms of Section~\ref{sec:2}. Then the selected kernel dictionary members within $R_a$ are divided into the different parallel-connected groups ($R_{a0}, R_{a1}, \cdots$) in an orderly fashion. Afterward, the online predictor within the first parallel-connected group $R_{a0}$ operates its weight vector updating procedure, and its prediction-error time series $TS_{a1}$ is continuously recorded. For the second parallel-connected group $R_{a1}$, the online predictor within $R_{a1}$ tracks the dynamic characteristics underlying the prediction-error time series $TS_{a1}$, and it makes a prediction for the forthcoming prediction errors in the prediction-error times series $TS_{a1}$. The online predictor within $R_{a1}$ is acquired by analyzing the effectiveness of different online algorithm candidates in the training procedure, and some of its parameters can be adjusted online with the sequentially arrived data. The aforementioned design principle for the online predictor within $R_{a1}$ also covers the online predictors in the following parallel-connected groups within $R_a$, and both the kernel dictionary selection and parallel connections for the first cascade-connected group $R_a$ has already been constructed. 

For the second cascade-connected group $R_b$, the online predictors within $R_b$ track the dynamic characteristics underlying the prediction-error time series $TS_{b0}$ that are recorded by its previous cascade-connected group $R_a$, and they make prediction for the forthcoming prediction errors in the prediction-error times series $TS_{b0}$. For the online predictors within $R_b$, the elements of the input vector may come from the recorded prediction-error time series $TS_{b0}$ or other variables provided by the previous cascade-connected groups, which leads to different kernel dictionary selection comparing with $R_a$. Also, the online predictors and parallel connections within $R_b$ are acquired by analyzing the effectiveness of different online algorithm candidates in the training procedure. In practice, the online predictors should be carefully constructed to achieve good prediction performance according to the effectiveness testings in the training procedure. 

The aforementioned design principle for the online predictor within $R_b$ also covers the online predictors in the following cascade-connected groups. As shown in Fig.~\ref{Connections}, the sum of all the generated online predictors, from the first cascade-connected group to a certain following cascade-connected group, constitutes the additive kernel model for the certain cascade-connected group. Most importantly, the adopted prediction function at a given time sample can be chosen from all the generated additive kernel models by online monitoring their prediction performances. The beneficial properties of the proposed approach and how specific modeling technique used affect its online prediction performance will be further studied in the next section.

In the online kernel modeling of nonstationary time series, intuitively the nonstationarity may be classified into two categories according to the prediction performance of the currently-adopted kernel model. One is the nonstationary trend, which can be captured by just optimizing the modeling parameters in the selected kernel model, and the other category is higher-order nonstationarity, which may need to be learned by changing the type of kernel function or even changing to other online prediction algorithms. The improved information interactions can enhance the currently-adopted kernel model's tracking ability as much as possible before the underlying nonstationarity is assigned to the second case, which may be beneficial to the selection procedure of online kernel models. Sometimes, if the unpredictable part of the nonstationarity has a great influence, the prediction performance can be a misleading indicator to determine whether the selected online kernel model should be changed. In this case, analyzing all the generated prediction-error time series can provide some useful clues to identify whether unpredictable nonstationarity accounts for a large part.

\section{Numerical Simulations}\label{sec:6}

Three nonstationary time series, a chaotic time series, the capacitor-current time series in the second-order circuit with the unstable equilibrium point, and the real-world sunspot time series, are chosen in the experiments to investigate the effectiveness of the proposed approach. For the KAF algorithms, the ALD-KRLS algorithm \cite{1315946}, the improved KRLS (IKRLS) algorithm \cite{9147041}\,\footnote{The correct weight-vector updating formula (11) in the reference \cite{9147041} is $\bm{\widetilde{\alpha}}(n) = \bm{\widetilde{\alpha}}(n-1)+\bm{Q}(n-1)\widetilde{\bm{k}}_{n-1}\big(y(n)-\widetilde{\bm{k}}_{n-1}^{\rm T}\widetilde{\bm{\alpha}}(n-1)\big)/\big(\lambda+\widetilde{\bm{k}}_{n-1}^{\rm T}\bm{Q}(n-1)\widetilde{\bm{k}}_{n-1}\big)$.} and the RRKOL algorithm \cite{7407373} are adopted as the representative algorithms to be studied. For the online tunable RBF algorithms, the fast tunable RBF (FT-RBF) algorithm \cite{7310878} and the fast tunable gradient RBF (FT-GRBF) algorithm~\cite{9040439} are adopted as the representative algorithms.

\subsection{Chaotic Time Series Online Prediction}\label{S6.1}

Derived from a finite mode truncation of the partial differential equations, the Lorenz time series\cite{lorenz1963deterministic,8291833} is given by three Lorenz differential equations
\begin{equation}\label{chaotic} 
\begin{aligned}
\frac{\mathrm{d}z_1(t)}{\mathrm{d}t} &= \sigma_1 \bigl( z_2(t) - z_1(t) \bigr) , \\
\frac{\mathrm{d}z_2(t)}{\mathrm{d}t} &= - z_1(t)z_3(t) + r z_1(t) - z_2(t) , \\
\frac{\mathrm{d}z_3(t)}{\mathrm{d}t} &= z_1(t)z_2(t) - b z_3(t) ,
\end{aligned}
\end{equation}
where $\sigma_1$, $r$ and $b$ are the parameters that control the behavior of the Lorenz system. The Lorenz time series samples are generated with the step size 0.01 and the starting point (0,1,0). The first $3000$ samples are set as the training dataset to obtain the optimal or appropriate parameters, and the $3000\sim8000$ samples are set as the testing dataset to examine the effectiveness of the studied algorithms. During the training procedure, the optimal or near-optimal parameter settings are acquired by validating the sequential prediction performance on the $2500\sim3000$ samples. For the online prediction of the dynamic input-output relationship in (\ref{chaotic}), we set the input vector as $\bm{x}_n = [z_1(n) ~ z_2(n) ~ z_3(n)]^{\rm T}$ to predict $y_{n}=z_2(n+5)$, with the time-varying control parameters $\sigma_1 = 10, b =\frac{1}{3} (4+3(1+\sin(0.1t))), r = 25+3\bigl( 1 + \cos \bigl( 2^{\rm {0.001t}}\bigr) \bigr)$ \cite{6334482,9040439}. Since the input vector $\bm{x}_n$ here does not explicitly contain the past output $y_n$, the differencing operation for the input of the FT-GRBF algorithm and the second recurrent term of (\ref{derivative}) in the RRKOL algorithm are not applied in this simulation \cite{9040439,7407373}. 

\begin{table}[htb!]\setcounter{table}{0}
\vspace*{-2mm}
\centering
\caption{Overview of representative algorithms from generalized optimization viewpoint} 
\vspace*{-1mm}
\label{overview} 
\begin{tabular}{cccc}
\toprule
Algorithm & $\widetilde{\bm{\Sigma}}^{-1}$ & Parallel Connections & Cascade Connections \\
\midrule
ALD-KRLS & $\surd$ & $\surd$ & $\surd$ \\
IKRLS & $\surd$ & $\surd$ & $\surd$ \\
RRKOL & $-$ & $-$ & $\surd$ \\
FT-RBF & $-$ & $-$ & $\surd$ \\
FT-GRBF & $-$ & $-$ & $\surd$ \\
\bottomrule
\end{tabular} 
\vspace*{-1mm}
\end{table} 

\begin{table*}[tp!]\setcounter{table}{1}
\vspace*{-2mm}
\centering
\caption{Prediction performance of the Lorenz time series with optimized $\widetilde{\bm{\Sigma}}^{-1}$ and generalized connections for IKRLS algorithm} 
\label{ikrlsprediction} 
\vspace*{-1mm}
\renewcommand\arraystretch{2.0}
\begin{tabular}{ccl|cccccccc}
\toprule
Algorithm & $m$ & $\widetilde{\bm{\Sigma}}^{-1}$ & Indicator & $R_a$ & $R_b$ & $R_c$ & $R_d$ & $R_e$ & $R_f$ & $R_g$ \\
\midrule
IKRLS & 6 & $V_1$ & \makecell{MAE \\ MSE} & $\makecell{0.092 \\ \bm{0.460}}$ & $\makecell{\bm{0.048} \\ 0.626}$ & $\makecell{0.066 \\ 1.081}$ & $\makecell{0.118 \\ 1.972}$ & $\makecell{0.229 \\ 3.686}$ & $\makecell{0.449 \\ 6.990}$ & $\makecell{0.890 \\ 13.380}$ \\ \hline
IKRLS & 6 & $V_2$ & \makecell{MAE \\ MSE} & $\makecell{0.054 \\ \bm{0.328}}$ & $\makecell{\bm{0.026} \\ 0.416}$ & $\makecell{0.030 \\ 0.680}$ & $\makecell{0.050 \\ 1.211}$ & $\makecell{0.093 \\ 2.236}$ & $\makecell{0.180 \\ 4.212}$ & $\makecell{0.353 \\ 8.027}$ \\ \hline
IKRLS & 3,3 & $V_2$ & \makecell{MAE \\ MSE} & $\makecell{0.270 \\ 0.381}$ & $\makecell{0.046 \\ 0.072}$ & $\makecell{0.011 \\ 0.022}$ & $\makecell{0.005 \\ 0.010}$ & $\makecell{\bm{0.003} \\ \bm{0.009}}$ & $\makecell{0.005 \\ 0.014}$ & $\makecell{0.008 \\ 0.025}$ \\ \hline
IKRLS & 3,3 & $Vp_1$ & \makecell{MAE \\ MSE} & $\makecell{0.084 \\ 0.131}$ & $\makecell{0.022 \\ 0.042}$ & $\makecell{0.010 \\ 0.023}$ & $\makecell{\bm{0.006} \\ \bm{0.018}}$ & $\makecell{0.006 \\ 0.019}$ & $\makecell{0.008 \\ 0.028}$ & $\makecell{0.014 \\ 0.047}$ \\ \hline
IKRLS & 1$\times$6 & $Vp_2$ & \makecell{MAE \\ MSE} & $\makecell{0.442 \\ 0.653}$ & $\makecell{0.124 \\ 0.306}$ & $\makecell{0.068 \\ \bm{0.238}}$ & $\makecell{\bm{0.053} \\ 0.254}$ & $\makecell{0.063 \\ 0.342}$ & $\makecell{0.090 \\ 0.528}$ & $\makecell{0.141 \\ 0.876}$ \\
\bottomrule
\end{tabular} 
\vspace*{-3mm}
\end{table*} 

\begin{table*}[b!]\setcounter{table}{3}
\vspace*{-4mm}
\centering
\caption{Prediction performance of the Lorenz time series with optimized $\widetilde{\bm{\Sigma}}^{-1}$ and generalized connections for ALD-KRLS algorithm} 
\label{aldprediction} 
\vspace*{-1mm}
\renewcommand\arraystretch{2.0}
\begin{tabular}{cclc|cccccccc}
\toprule
Algorithm & $m$ & $D$ & $\widetilde{\bm{\Sigma}}^{-1}$ & Indicator & $R_a$ & $R_b$ & $R_c$ & $R_d$ & $R_e$ & $R_f$ & $R_g$ \\
\midrule
ALD-KRLS & 13 & $D_{ald}$ & $-$ & \makecell{MAE \\ MSE} & $\makecell{0.268 \\ 0.435}$ & $\makecell{0.029 \\ 0.060}$ & $\makecell{0.007 \\ 0.018}$ & $\makecell{0.003 \\ 0.007}$ & $\makecell{\bm{0.002} \\ \bm{0.006}}$ & $\makecell{0.003 \\ 0.009}$ & $\makecell{0.006 \\ 0.017}$ \\ \hline
ALD-KRLS & 6 & $D'_{ald}$ & $-$ & \makecell{MAE \\ MSE} & $\makecell{0.660 \\ 0.999}$ & $\makecell{0.062 \\ 0.107}$ & $\makecell{0.013 \\ 0.026}$ & $\makecell{0.004 \\ 0.010}$ & $\makecell{\bm{0.003} \\ \bm{0.007}}$ & $\makecell{0.003 \\ 0.010}$ & $\makecell{0.006 \\ 0.017}$ \\ \hline
ALD-KRLS & 6 & $D_m$ & $-$ & \makecell{MAE \\ MSE} & $\makecell{0.649 \\ 1.001}$ & $\makecell{0.061 \\ 0.107}$ & $\makecell{0.012 \\ 0.025}$ & $\makecell{0.004 \\ 0.010}$ & $\makecell{\bm{0.003} \\ \bm{0.007}}$ & $\makecell{0.003 \\ 0.010}$ & $\makecell{0.006 \\ 0.017}$ \\ \hline
ALD-KRLS & 6 & $D_m$ & $\surd$ & \makecell{MAE \\ MSE} & $\makecell{0.313 \\ 0.414}$ & $\makecell{0.030 \\ 0.049}$ & $\makecell{0.007 \\ 0.015}$ & $\makecell{0.003 \\ 0.007}$ & $\makecell{\bm{0.002} \\ \bm{0.006}}$ & $\makecell{0.003 \\ 0.009}$ & $\makecell{0.005 \\ 0.017}$ \\ 
\bottomrule
\end{tabular} 
\vspace*{-1mm}
\end{table*} 

An overview of the five representative algorithms from the enhanced information interaction viewpoint is summarized in Table~\ref{overview}. Owing to the tunable structures of the online tunable RBF/GRBF algorithms and the lack of kernel dictionary selection procedure in the RRKOL algorithm, only the ALD-KRLS and IKRLS algorithms are adopted to fully examine the effectiveness of the improved information interactions in this simulation. However, all the five algorithms with cascade connections are compared to reveal the importance of the clues in underlying prediction-error time series. We start the selection procedure for the isotropic form of kernel covariance matrix in (\ref{optipredictor}), namely, in the form of kernel bandwidth, and the following experiments are with the acquired isotropic matrices as the initial kernel covariance matrices.

For the IKRLS algorithm, the prediction performance indicators for the test dataset of $3000\sim8000$ samples are shown in Table~\ref{ikrlsprediction}. In the first cascade-connected group $R_a$, the selected kernel dictionary $D_m$ consists of $m=6$ members. The $D_m$ can be divided into different groups in parallel connections, such as (3,3) which denotes that the first parallel-connected group includes 3 kernel regressors and the second parallel-connected group also includes 3 kernel regressors, and ($1\times 6$) which denotes that there are a total of 6 parallel-connected groups and each parallel-connected group includes one kernel regressor. The optimized $V_1$ is the acquired isotropic kernel covariance matrix, $V_2$ is the optimized symmetric kernel covariance matrix with no parallel connections, while $Vp_1$ is the set of optimized symmetric kernel covariance matrices that come from their respective parallel-connected groups, and so is $Vp_2$. The kernel dictionary selection criteria in the IKRLS algorithm include both the ALD criterion and the distance criterion\cite{9147041}, and thus Algorithm~\ref{aldcma} is adopted to optimize $\widetilde{\bm{\Sigma}}^{-1}$. $R_a, R_b,\cdots ,R_g$ denote the cascade-connected groups as illustrated in Fig.~\ref{Connections}. As described in the training procedure of Algorithm~\ref{completeness}, the kernel dictionary selection and parallel connections for the first cascade-connected group $R_a$ has already been constructed. The respective key parameters are described in the left side of the vertical line in Table~\ref{ikrlsprediction}, and the corresponding MAE/MSE metrics for the online prediction procedure are listed under $R_a$.

For the second cascade-connected group $R_b$, if the input vector is explicitly composed of the past signals in the prediction-error time series recorded by $R_a$, then the simplest online predictor is using the last datum in the recorded prediction-error time series to predict the current forthcoming prediction error, i.e., for the previous cascade-connected group $R_a$, setting the latest prediction error as the current error compensation for its online predictor. If the simplest online predictors are applied to $R_b$ and all the following cascade-connected groups, then to a certain extent, these online predictors actually capture the high-order gradients in the prediction-error time series recorded by $R_a$. For more information about the high-order gradients, the reader is referred to a typical RBF model in \cite{478403}. In this simulation, after analyzing the prediction-error time series that is recorded by $R_a$ in the training procedure, we find that capturing the high-order gradients in the prediction-error time series is more efficient than continuously using the Gaussian kernel function modeling. Thus in the following cascade-connected groups, we use the simplest online predictors to show the effectiveness of the proposed information interactions, where the kernel dictionary selection procedures are not performed. Note that some improved online predictors can be applied to obtain better prediction performance in practice, such as the linear recursive least squares algorithm and online kernel algorithms. As shown in Table~\ref{ikrlsprediction}, the corresponding MAE/MSE metrics for each cascade-connected group in the online prediction procedure are listed under $R_b, \cdots , R_g$, respectively.

How the connection modes and key parameters in the first cascade-connected group affect the prediction performances in the following cascade-connected groups can be revealed. For example, it can be seen that the best prediction performance is attained at the $R_d$/$R_e$ cascade-connected group with the (3,3) parallel-connected first cascade-connected group. The results of Table~\ref{ikrlsprediction} also indicate that the cascade connections can help to capture the underlying dynamic characteristics in the prediction-error time series. By comparing the MAE/MSE metrics of the first two IKRLS predictors with those of the last three IKRLS predictors in Table~\ref{ikrlsprediction}, we can clearly observe the important role of the improved information interactions in the key-parameters selection for the IKRLS algorithm.

\begin{table}[htb!]\setcounter{table}{2}
\vspace*{-2mm}
\centering
\caption{Optimization of $\widetilde{\bm{\Sigma}}^{-1}$ with different forms for IKRLS} 
\vspace*{-2mm}
\label{symdiag} 
\begin{tabular}{cclcc}
\toprule
Algorithm & $m$ & $\widetilde{\bm{\Sigma}}^{-1}$ & MAE & MSE \\
\midrule
IKRLS & 6 & $V_1$ & 0.092 & 0.460 \\
IKRLS & 6 & $V_{21}$ & 0.065 & 0.354 \\
IKRLS & 6 & $V_{22}$ & 0.055 & 0.330 \\
IKRLS & 6 & $V_2$ & 0.054 & 0.328 \\
IKRLS & 6 & $Vd_{21}$ & 0.163 & 2.334 \\
IKRLS & 6 & $Vd_{22}$ & 0.060 & 0.155 \\
IKRLS & 6 & $Vd_{23}$ & 0.572 & 16.945 \\
IKRLS & 6 & $Vd_2$ & 0.074 & 0.638 \\
\bottomrule
\end{tabular} 
\vspace*{1mm}
\end{table}

\begin{table*}[b!]\setcounter{table}{4}
\vspace*{-5mm}
\centering
\caption{Prediction performance of the Lorenz time series with cascade connections for various algorithms} 
\label{chaoticprediction} 
\vspace*{-1mm}
\renewcommand\arraystretch{2.0}
\begin{tabular}{ccc|cccccccc}
\toprule
Algorithm & $m$ & $\beta$ & Indicator & $R_a$ & $R_b$ & $R_c$ & $R_d$ & $R_e$ & $R_f$ & $R_g$ \\
\midrule
ALD-KRLS & 13 & $ - $ & \makecell{MAE \\ MSE} & $\makecell{0.268 \\ 0.435}$ & $\makecell{0.029 \\ 0.060}$ & $\makecell{0.007 \\ 0.018}$ & $\makecell{0.003 \\ 0.007}$ & $\makecell{\bm{0.002} \\ \bm{0.006}}$ & $\makecell{0.003 \\ 0.009}$ & $\makecell{0.006 \\ 0.017}$ \\ \hline
IKRLS & 6 & $0.90$ & \makecell{MAE \\ MSE} & $\makecell{0.092 \\ \bm{0.460}}$ & $\makecell{\bm{0.048} \\ 0.626}$ & $\makecell{0.066 \\ 1.081}$ & $\makecell{0.118 \\ 1.972}$ & $\makecell{0.229 \\ 3.686}$ & $\makecell{0.449 \\ 6.990}$ & $\makecell{0.890 \\ 13.380}$ \\ \hline
RRKOL & 6 & $ - $ & \makecell{MAE \\ MSE} & $\makecell{0.440 \\ 0.647}$ & $\makecell{0.056 \\ 0.096}$ & $\makecell{0.011 \\ 0.021}$ & $\makecell{\bm{0.004} \\ \bm{0.008}}$ & $\makecell{0.004 \\ 0.011}$ & $\makecell{0.008 \\ 0.021}$ & $\makecell{0.015 \\ 0.041}$ \\ \hline
FT-RBF & 10 & $0.70$ & \makecell{MAE \\ MSE} & $\makecell{\bm{0.046} \\ \bm{0.352}}$ & $\makecell{0.048 \\ 0.497}$ & $\makecell{0.081 \\ 0.865}$ & $\makecell{0.154 \\ 1.586}$ & $\makecell{0.297 \\ 2.974}$ & $\makecell{0.583 \\ 5.655}$ & $\makecell{1.145 \\ 10.843}$ \\ \hline
FT-GRBF & 6 & $0.90$ & \makecell{MAE \\ MSE} & $\makecell{0.023 \\ 0.036}$ & $\makecell{0.007 \\ \bm{0.017}}$ & $\makecell{\bm{0.006} \\ 0.022}$ & $\makecell{0.008 \\ 0.038}$ & $\makecell{0.015 \\ 0.068}$ & $\makecell{0.030 \\ 0.127}$ & $\makecell{0.058 \\ 0.241}$ \\ 
\bottomrule
\end{tabular} 
\vspace*{-2mm}
\end{table*}

In Table~\ref{symdiag}, the isotropic kernel covariance matrix $V_1$ is iteratively optimized to the diagonal form $Vd_2$ and to the symmetric form $V_2$, using Algorithm~\ref{aldcma}. As expected, the online prediction performance is enhanced with the symmetric general kernel covariance matrix. More specifically, the respective evolution paths are $V_1 \rightarrow V_{21} \rightarrow V_{22} \rightarrow V_{2}$ and $V_1 \rightarrow Vd_{21} \rightarrow Vd_{22} \rightarrow Vd_{23} \rightarrow Vd_{2}$, and the progressively enhanced prediction performances in each evolution path are validated in the training procedure. Comparing the MAE/MSE metrics in each evolution path given in Table~\ref{symdiag}, we can observe that optimizing the isotropic kernel covariance matrix with (\ref{optiform}) behaves stably and robustly in the online prediction procedure, which implies that the improved flexibility in kernel structure helps to enhance the generalization ability.

In Table~\ref{aldprediction}, the ALD-KRLS algorithm is operated with the kernel dictionary $D_m$, the ALD criterion based kernel dictionary $D_{ald}$ with 13 members, and the ALD criterion based kernel dictionary $D'_{ald}$ with 6 members, respectively. It can be observed that the last ALD-KRLS with the optimized symmetric $\widetilde{\bm{\Sigma}}^{-1}$ achieves the same prediction performance with much smaller kernel dictionary size than the first ALD-KRLS with the isotropic kernel matrices $D_{ald}$. The last ALD-KRLS also obtains a better prediction performance than the second and third ALD-KRLS algorithms. This implies that the optimized symmetric form of $\widetilde{\bm{\Sigma}}^{-1}$ can alleviate the prediction uncertainty caused by the selected kernel dictionaries. 

\begin{figure}[htb!]
\vspace*{-4mm}
\begin{center}
\includegraphics[width=\columnwidth]{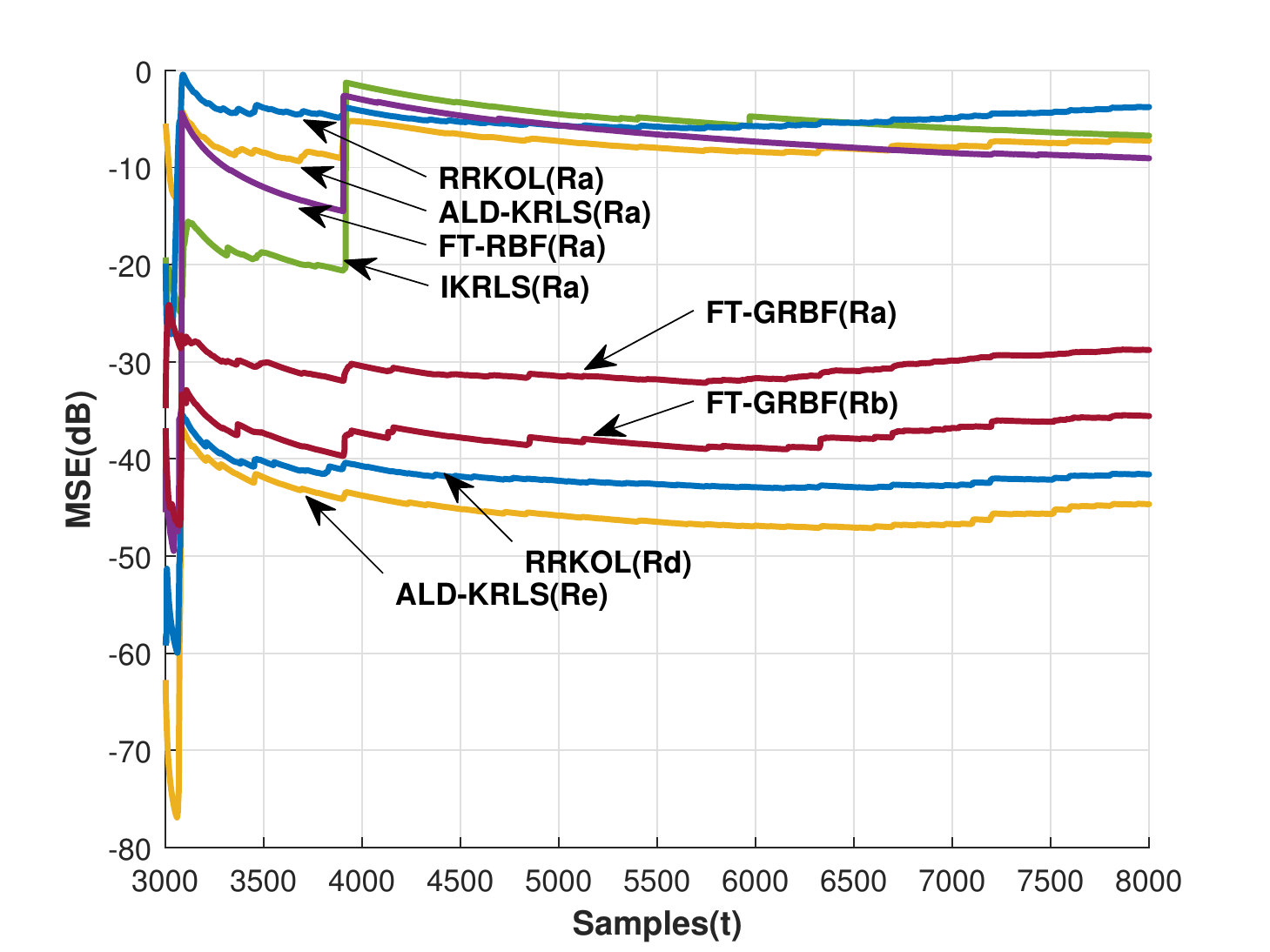}
\end{center}
\vspace*{-6mm}
\caption{Prediction performance comparison of the Lorenz time series with cascade connections for various algorithms.} 
\label{chaoticfig} 
\vspace*{-1mm}
\end{figure}

Table~\ref{chaoticprediction} and Fig.~\ref{chaoticfig} compare the prediction performances of all the five algorithms with cascade connections. Observe that except for the IKRLS and FT-RBF, the optimal prediction performances (marked in bold) are significantly better than the performances of the first cascade-connected groups $R_a$, which indicates that the improved information interaction helps to select superior online kernel algorithms.

\begin{table*}[tb!]\setcounter{table}{5}
\vspace*{-2mm}
\centering
\caption{Prediction performance of the capacitor-current time series with generalized connections for various algorithms} 
\label{circuitprediction} 
\vspace*{-1mm}
\renewcommand\arraystretch{2.0}
\begin{tabular}{cccc|cccccccc}
\toprule
Algorithms & $m$ & $\mathnormal{h}_{0}$ & $\beta$ & Indicators & $R_a$ & $R_b$ & $R_c$ & $R_d$ & $R_e$ & $R_f$ & $R_g$ \\
\midrule
ALD-KRLS & 7 & $1e10$ & $-$ & \makecell{MAE \\ MSE} & $\makecell{1.314 \\ 3.021}$ & $\makecell{0.029 \\ 0.061}$ & $\makecell{\bm{0.002} \\ \bm{0.004}}$ & $\makecell{0.002 \\ 0.006}$ & $\makecell{0.004 \\ 0.012}$ & $\makecell{0.007 \\ 0.022}$ & $\makecell{0.014 \\ 0.041}$ \\ \hline
ALD-KRLS & 4,3 & $2e10$ & $-$ & \makecell{MAE \\ MSE} & $\makecell{0.093 \\ 0.316}$ & $\makecell{0.005 \\ 0.021}$ & $\makecell{\bm{0.001} \\ \bm{0.004}}$ & $\makecell{0.002 \\ 0.007}$ & $\makecell{0.004 \\ 0.012}$ & $\makecell{0.007 \\ 0.022}$ & $\makecell{0.014 \\ 0.042}$ \\ \hline
IKRLS & 6 & $2e10$ & $0.005$ & \makecell{MAE \\ MSE} & $\makecell{2.454 \\ 4.925}$ & $\makecell{0.093 \\ \bm{0.468}}$ & $\makecell{\bm{0.060} \\ 0.782}$ & $\makecell{0.118 \\ 1.427}$ & $\makecell{0.236 \\ 2.669}$ & $\makecell{0.473 \\ 5.063}$ & $\makecell{0.945 \\ 9.693}$ \\ \hline
RRKOL & 7 & $2e10$ & $-$ & \makecell{MAE \\ MSE} & $\makecell{10.388 \\ 21.128}$ & $\makecell{0.266 \\ 0.504}$ & $\makecell{0.007 \\ 0.012}$ & $\makecell{\bm{0.002} \\ \bm{0.006}}$ & $\makecell{0.003 \\ 0.011}$ & $\makecell{0.006 \\ 0.020}$ & $\makecell{0.013 \\ 0.039}$ \\  \hline
RRKOL & 4,3 & $2e10$ & $-$ & \makecell{MAE \\ MSE} & $\makecell{0.678 \\ 1.287}$ & $\makecell{0.017 \\ 0.029}$ & $\makecell{\bm{0.002} \\ \bm{0.004}}$ & $\makecell{0.002 \\ 0.007}$ & $\makecell{0.005 \\ 0.013}$ & $\makecell{0.009 \\ 0.025}$ & $\makecell{0.018 \\ 0.048}$ \\ \hline
FT-RBF & 10 & $1e9$ & $0.05$ & \makecell{MAE \\ MSE} & $\makecell{1.095 \\ 1.757}$ & $\makecell{\bm{0.220} \\ \bm{0.596}}$ & $\makecell{0.362 \\ 1.010}$ & $\makecell{0.658 \\ 1.842}$ & $\makecell{1.239 \\ 3.449}$ & $\makecell{2.349 \\ 6.550}$ & $\makecell{4.509 \\ 12.552}$ \\ \hline
FT-GRBF & 7 & $1e10$ & $0.01$ & \makecell{MAE \\ MSE} & $\makecell{\bm{0.043} \\ \bm{0.100}}$ & $\makecell{0.054 \\ 0.143}$ & $\makecell{0.095 \\ 0.252}$ & $\makecell{0.176 \\ 0.464}$ & $\makecell{0.333 \\ 0.872}$ & $\makecell{0.640 \\ 1.663}$ & $\makecell{1.236 \\ 3.197}$ \\ 
\bottomrule
\end{tabular} 
\vspace*{-4mm}
\end{table*}

\subsection{Online Prediction of the Capacitor-Current Time Series}\label{S6.2}

For this second-order \emph{RLC} circuit, the relationship between the capacitor-current $x_{c1}(t)$ and the capacitor-voltage $x_{c2}(t)$ is nonstationary with the unstable equilibrium point \cite{fundamentals}, which can be described as,
\begin{equation}\label{rlc} 
\begin{aligned}
\frac{\mathrm{d}x_{c1}(t)}{\mathrm{d}t} &= x_{c2}(t) , \\
\frac{\mathrm{d}x_{c2}(t)}{\mathrm{d}t} &= -\omega_c^{2}(t)x_{c1}(t) - 2\delta_{t}x_{c2}(t) ,
\end{aligned}
\end{equation}
where the control parameters are set as $\omega_c(t) = 5\cos(0.05t)$ and $\delta_{t} = -\frac{1}{2}$ to enhance the nonstationarity. The simulated samples are generated with the step size of 0.008 and the starting point (0,0.30). The first $500$ samples are used as the training dataset, and the $500\sim2500$ samples are set as the testing dataset. In the training procedure, the optimal or appropriate parameter settings can be acquired by validating the sequential prediction performance on the first $300\sim500$ samples. We set the input vector as $\bm{x}_n =\big[x_{c1}(n) ~ x_{c2}(n)\big]^{\rm T}$ to predict $y_n=x_{c2}(n+1)$, i.e., this is a one-step-ahead prediction. Again, the differencing operation for the input of the FT-GRBF algorithm is not applied, but the second recurrent term of (\ref{derivative}) in the RRKOL algorithm is applied since the input $\bm{x}_n$ contains the past output $y_{n-1}=x_{c2}(n)$. The adopted optimization strategy in this simulation is the same as in Subsection~\ref{S6.1}.

\begin{figure}[htb!]
\vspace*{-3mm}
\begin{center}
\includegraphics[width=\columnwidth]{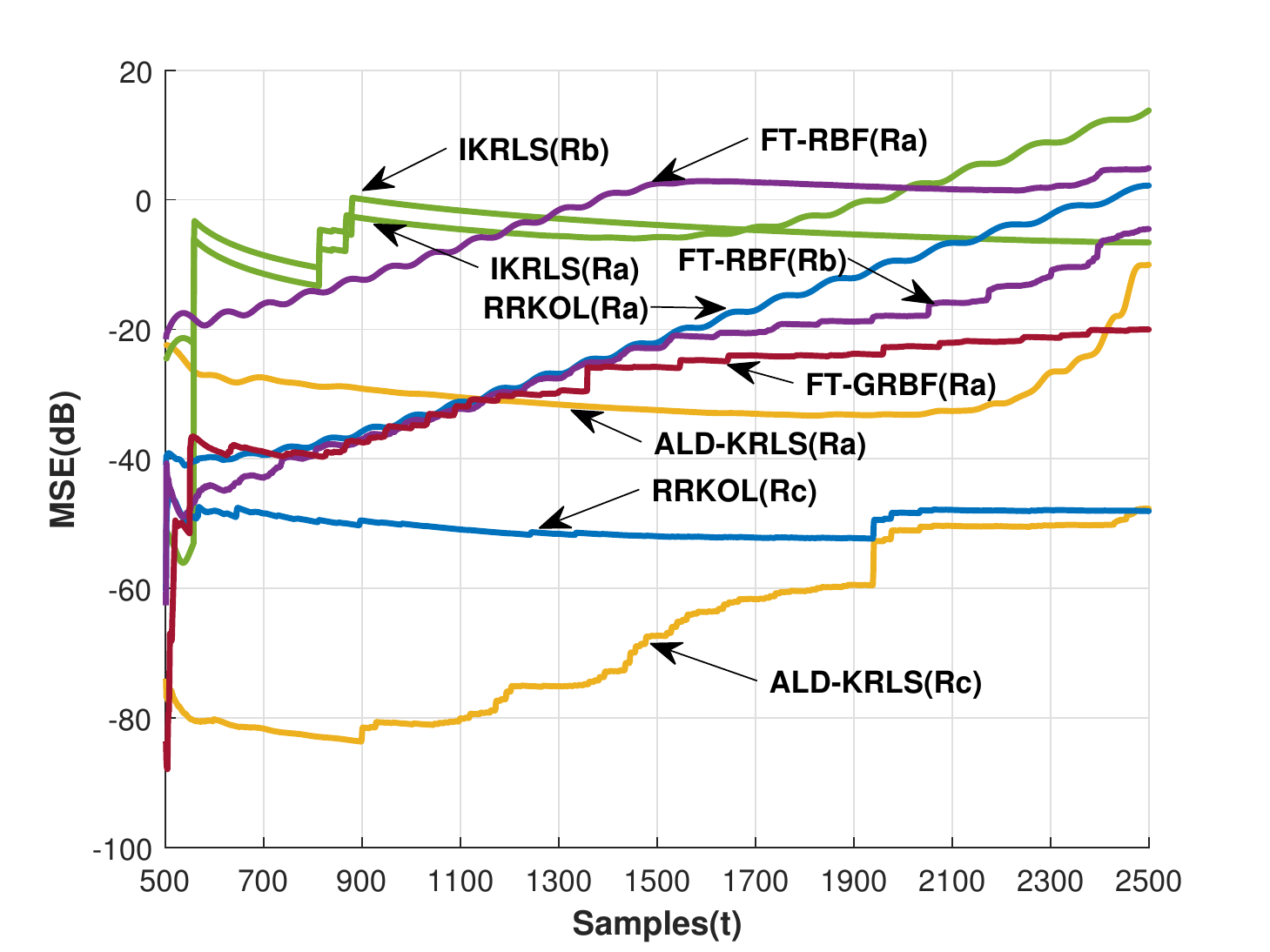}
\end{center}
\vspace*{-6mm}
\caption{Prediction performance comparison of the capacitor-current time series with generalized connections for various algorithms.} 
\label{circuitfig} 
\vspace*{-1mm}
\end{figure}

Table \ref{circuitprediction} and Fig. \ref{circuitfig} compare the prediction performances of the representative algorithms with both parallel connections and cascade connections. The cascade connections in each algorithm (except for the FT-GRBF algorithm) obtain better prediction performances in the following cascade-connected groups, which indicates that the cascade connections can help to capture the underlying dynamic characteristics in the prediction-error time series. For the ALD-KRLS and RROKL algorithms, their respective parallel-connected cases obtain better prediction performances, which indicates that the parallel connections also can help to capture the underlying dynamic characteristics in the prediction-error time series. Therefore, the effectiveness of the generalized kernel connections is demonstrated in this simulation. The online tunable RBF algorithms can obtain better prediction performances than the KAF algorithms in the first cascade-connected group, but this is not the case for the optimal prediction performances among cascade-connected groups.

\begin{table}[bp!]
\vspace*{-5mm}
\centering
\caption{Prediction performance of the sunspot time series with Cascade Connections for various algorithms} 
\label{sunspotprediction}
\vspace*{-2mm}
\renewcommand\arraystretch{2.0}
\begin{tabular}{cccc|ccc}
\toprule
Algorithms & $m$ & $p_r$ & $\beta_2$ & Indicators & $R_a$ & $R_b$ \\
\midrule
ALD-KRLS & 7 & 1 & $0.96$ & \makecell{MAE \\ MSE} & $\makecell{6.451 \\ 12.036}$ & $\makecell{1.525 \\ 2.185}$ \\ \hline
IKRLS & 5 & 11 & $0.98$ & \makecell{MAE \\ MSE} & $\makecell{2.204 \\ 3.520}$ & $\makecell{2.097 \\ 3.356}$ \\ \hline
RRKOL & 7 & 17 & $0.98$ & \makecell{MAE \\ MSE} & $\makecell{9.607 \\ 15.776}$ & $\makecell{1.422 \\ 2.082}$ \\ \hline
FT-RBF & 23 & 33 & $0.92$ & \makecell{MAE \\ MSE} & $\makecell{2.143 \\ 2.990}$ & $\makecell{1.730 \\ 2.612}$ \\ \hline
FT-GRBF & 12 & 11 & $0.98$ & \makecell{MAE \\ MSE} & $\makecell{2.386 \\ 4.566}$ & $\makecell{2.119 \\ 3.950}$ \\ 
\bottomrule
\end{tabular} 
\vspace*{-1mm}
\end{table}

\subsection{Sunspot Time Series Online Prediction}\label{S6.3}

The sunspot time series is composed of annual averaged numbers of observed sunspots, which is a widely used benchmark that contains nonstationarity~\cite{9040439}. The one-step ahead prediction of the monthly recorded sunspot time series $\{x_s(n)\}$ from 1830 to 2019 is considered. The input vector of the predictor is set to $\bm{x}_n\! =\! \big[x_s(n-1) ~ x_s(n-2) ~ x_s(n-3) ~ x_s(n-4)]^{\rm T}$ and the desired output is $y_n\! =\! x_s(n)$ in this simulation. The first 500 samples are used as the training dataset, and the $500\sim2280$ samples are set as the testing dataset to examine the effectiveness of the studied algorithms. In the training procedure, the optimal or appropriate parameter settings can be acquired by validating the sequential prediction performance on the first $300\sim500$ samples. Since the input vector is composed of the past output signals of $y_n$, both the differencing procedure in the FT-GRBF algorithm and the second recurrent term of (\ref{derivative}) in the RRKOL algorithm can be applied in this simulation. As the sunspot time series is monthly recorded and annually averaged, there exists strong linearity among the neighboring time series samples. The cascade connections can provide a structure to combine the kernel based online modeling approaches with linear recursive least squares algorithm in this simulation, i.e., the first cascade-connected group adopts the five representative algorithms, and the second cascade-connected group adopts the linear recursive least squares algorithm. 

\begin{figure}[bp!]
\vspace*{-10mm}
\begin{center}
\includegraphics[width=\columnwidth]{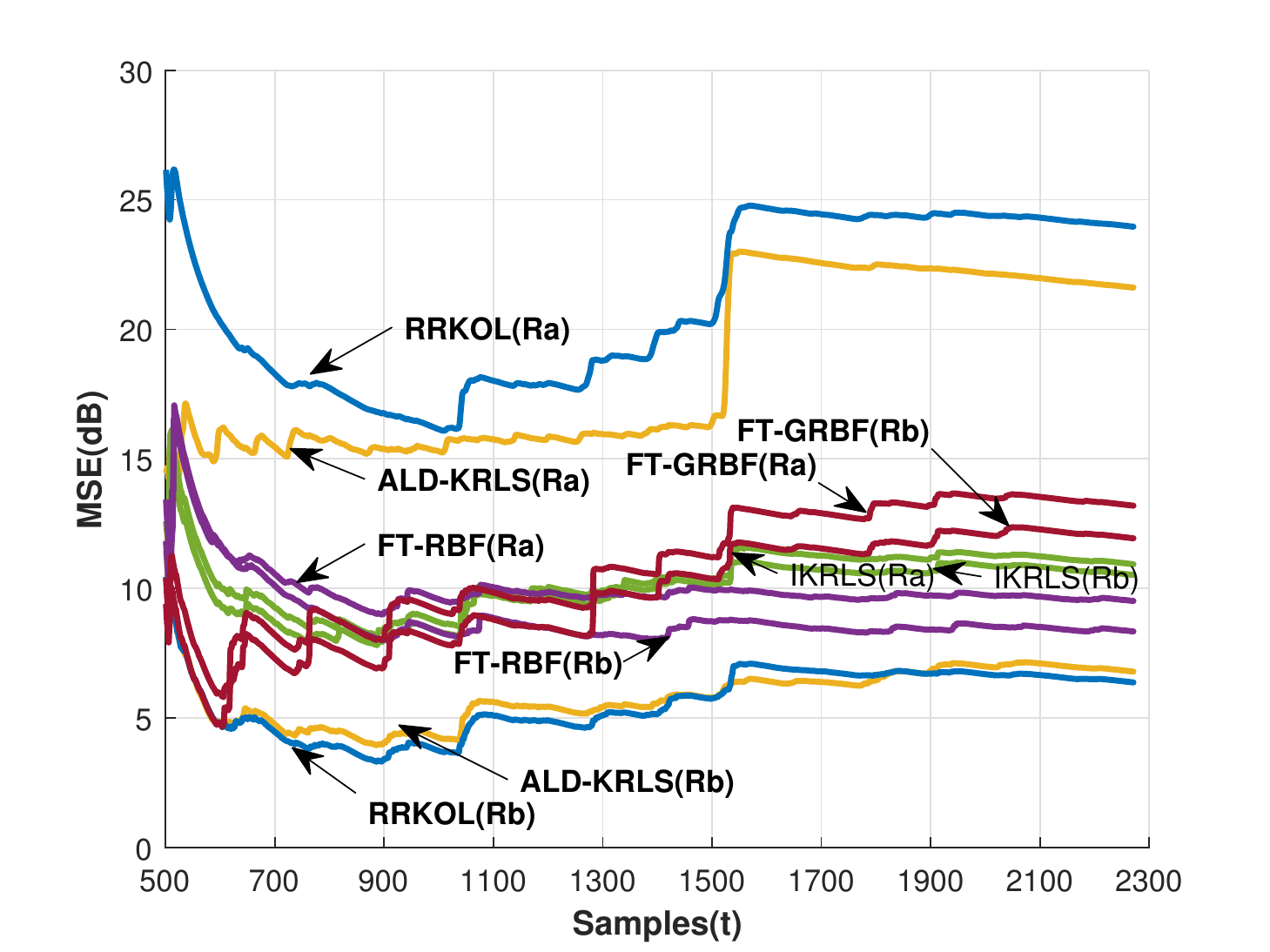}
\end{center}
\vspace*{-6mm}
\caption{Prediction performance of the sunspot time series with cascade connections for various algorithms.} 
\label{sunspotfig} 
\vspace*{-1mm}
\end{figure}

Table~\ref{sunspotprediction} and Fig.~\ref{sunspotfig} compare the prediction performances of the representative algorithms with cascade connections, where $p_r$ and $\beta_2$ respectively denote the dimension of the input vector and the exponential forgetting factor, in the linear recursive least squares algorithm. Each representative algorithm obtains better prediction performance in the second cascade connected group, especially for the ALD-KRLS and RRKOL algorithms. This demonstrate the effectiveness of the cascade connections in terms of providing a structure to combine complementary algorithms.

\section{Conclusions}\label{sec:7}

In this paper, we have proposed a structure parameter optimized kernel based online prediction approach with a generalized optimization strategy for nonstationary time series. The intermittent optimization of the real symmetric kernel covariance matrix has been realized to improve the kernel structure's flexibility and to alleviate the prediction uncertainty caused by the kernel dictionary selection procedure for nonstationary data. A generalized optimization strategy with multiple kernel connection modes has been designed to provide a self-contained way for constructing the entire connections of kernel regressors, with the enhanced ability to track the changing dynamic characteristics. The improved information interaction not only enhances the ability to deal with various online modeling problems but also provides a more flexible kernel structure and a more compatible way of kernel connections that can be combined with other existing online modeling techniques. 

\bibliographystyle{IEEEtran}

%



\end{document}